\documentclass[journal]{IEEEtran}
\IEEEoverridecommandlockouts
\usepackage{cite}
\usepackage{amsmath,amssymb,amsfonts}
\usepackage{graphicx}
\usepackage{textcomp}
\usepackage{xcolor}
\usepackage{algorithm}
\usepackage{algorithmic}
\usepackage{enumitem}
\usepackage{footnote}
\usepackage{multirow}
\usepackage{threeparttable}
\usepackage[colorlinks,linkcolor=blue]{hyperref}
\ifCLASSOPTIONcompsoc
\usepackage[caption=false, font=normalsize, labelfont=sf, textfont=sf]{subfig}
\else
\usepackage[caption=false, font=footnotesize]{subfig}
\fi
\usepackage{array}
\usepackage{scalerel}
\usepackage{siunitx,upgreek}

\def\BibTeX{{\rm B\kern-.05em{\sc i\kern-.025em b}\kern-.08em
    T\kern-.1667em\lower.7ex\hbox{E}\kern-.125emX}}

\usepackage[justification=centering]{caption}

\usepackage{etoolbox}
\let\mybibitem\bibitem
\renewcommand{\bibitem}[1]{%
  \ifstrequal{#1}{nature}
    {\color{blue}\mybibitem{#1}}
    {\color{black}\mybibitem{#1}}%
}
\begin{document}

\title{Pay Less But Get More: A Dual-Attention-based Channel Estimation Network for Massive MIMO Systems with Low-Density Pilots}

\author{Binggui Zhou,
        Xi Yang,
        Shaodan Ma,
        Feifei Gao,
        and Guanghua Yang
\thanks{This work was supported in part by the National Natural Science Foundation of China under Grants 62171201, 62261160650, and 62301221; in part by the Major Talent Program of Guangdong Provincial under Grant 2019QN01S103; in part by the Shanghai Pujiang Program under Grant 22PJ1403100; in part by the Science and Technology Development Fund, Macau SAR, under Grants 0087/2022/AFJ and SKL-IOTSC(UM)-2021-2023; in part by the Research Committee of University of Macau under Grant MYRG2020-00095-FST; and in part by the Beijing Natural Science Foundation under Grant L222002. (Corresponding authors: Shaodan Ma; Xi Yang.)}
\thanks{Binggui Zhou is with the School of Intelligent Systems Science and Engineering, Jinan University, Zhuhai 519070, China; and also with the State Key Laboratory of Internet of Things for Smart City and the Department of Electrical and Computer Engineering, University of Macau, Macao 999078, China (e-mail: binggui.zhou@connect.um.edu.mo).}
\thanks{Xi Yang is with the Shanghai Key Laboratory of Multidimensional Information Processing, School of Communication and Electronic Engineering, East China Normal University, Shanghai 200241, China (e-mail: xyang@cee.ecnu.edu.cn).}
\thanks{Shaodan Ma is with the State Key Laboratory of Internet of Things for Smart City and the Department of Electrical and Computer Engineering, University of Macau, Macao 999078, China (e-mail: shaodanma@um.edu.mo).}
\thanks{Feifei Gao is with the Institute for Artificial Intelligence, Tsinghua University (THUAI), State Key Lab of Intelligent Technologies and Systems, Tsinghua University, Beijing National Research Center for Information Science and Technology (BNRist), and Department of Automation, Tsinghua University, Beijing 100084, China (email: feifeigao@ieee.org).}
\thanks{Guanghua Yang is with the School of Intelligent Systems Science and Engineering and the GBA and B\&R International Joint Research Center for Smart Logistics, Jinan University, Zhuhai 519070, China (e-mail: ghyang@jnu.edu.cn).}
\thanks{The source code of this work is available at \href{https://github.com/bgzhou/DACEN/}{https://github.com/bgzhou/DACEN/}.}
}

\markboth{IEEE Transactions on Wireless Communications}{Zhou \MakeLowercase{\textit{et al.}}: Pay Less But Get More: A Dual-Attention-based Channel Estimation Network for Massive MIMO Systems with Low-Density Pilots}

\maketitle

\IEEEpubid{\begin{minipage}[t]{\textwidth}\ \\[12pt] \centering
    \copyright \ 2023 IEEE. Personal use of this material is permitted. Permission from IEEE must be obtained for all other uses, in any current or future media, including reprinting/republishing this material for advertising or promotional purposes, creating new collective works, for resale or redistribution to servers or lists, or reuse of any copyrighted component of this work in other works.
  \end{minipage}} 

\begin{abstract}

To reap the promising benefits of massive multiple-input multiple-output (MIMO) systems, accurate channel state information (CSI) is required through channel estimation. However, due to the complicated wireless propagation environment and large-scale antenna arrays, precise channel estimation for massive MIMO systems is significantly challenging and costs an enormous training overhead. Considerable time-frequency resources are consumed to acquire sufficient accuracy of CSI, which thus severely degrades systems' spectral and energy efficiencies. In this paper, we propose a dual-attention-based channel estimation network (DACEN) to realize accurate channel estimation via low-density pilots, by jointly learning the spatial-temporal domain features of massive MIMO channels with the temporal attention module and the spatial attention module. To further improve the estimation accuracy, we propose a parameter-instance transfer learning approach to transfer the channel knowledge learned from the high-density pilots pre-acquired during the training dataset collection period. Experimental results reveal that the proposed DACEN-based method achieves better channel estimation performance than the existing methods under various pilot-density settings and signal-to-noise ratios. Additionally, with the proposed parameter-instance transfer learning approach, the DACEN-based method achieves additional performance gain, thereby further demonstrating the effectiveness and superiority of the proposed method.

\end{abstract}

\begin{IEEEkeywords}
Low-overhead Channel Estimation, Massive MIMO, Attention Mechanism, Transfer Learning, Deep Learning
\end{IEEEkeywords}

\section{Introduction}

\IEEEPARstart MASSIVE multiple-input multiple-output (MIMO) has been the fundamental technology of the fifth-generation (5G) wireless communication systems and will continue to play an important role in the sixth-generation (6G) \cite{marzetta2015massive}. Although the large-scale antenna array of massive MIMO systems enables large spatial degrees of freedom, diversity, and array gains, which greatly improve the spectral and energy efficiencies of wireless systems \cite{larsson2014massive}, channel state information (CSI) is needed to be accurately acquired through channel estimation to reap these promising benefits.

Channel estimation for massive MIMO systems can be executed by inserting pilot symbols into the transmitted signal and then estimating the channels based on the received signal. Accurate channel estimates can be attained once we can afford the pilot training overhead and sacrifice sufficient time-frequency resources to the pilot symbols.\footnote{Although both channel state information reference signal (CSI-RS) and demodulation reference signal (DM-RS) can be regarded as pilots, in this paper, we mainly focus on CSI-RS.} The methods to conduct channel estimation can be roughly classified as the traditional estimation methods and the deep learning-based methods. Traditional estimation methods \cite{nguyen-le2010pilotaided,wang2005robust,chen2002maximum,geng2012dualdiagonal, huang2007lowcomplexity} estimate channels mainly based on the assumption of specific channel models or channel distributions. Hence, it is necessary for traditional methods to have a proper channel model or know prior information on channel distribution. With the increasing antenna array scale, the pilot training overhead and the computational complexity of traditional methods also increase significantly due to the enlarged channel dimension. To support practical massive MIMO applications, accurate channel estimation but with low pilot overhead is highly desirable and has attracted considerable investigation recently. Existing low-overhead channel estimation methods are primarily based on the sparsity property of massive MIMO channels. Their underlying principle is that, with channel sparsity, the CSI estimation can be transformed to the estimation of channel parameters with a significantly reduced number, e.g., the angle-of-arrivals (AoAs), the angle-of-departures (AoDs), which thereby facilitates low-overhead channel estimation. For example, many studies have employed compressive sensing (CS), e.g., the orthogonal matching pursuit (OMP) algorithm \cite{lee2016channel, wu2019lowcomplexity}, to estimate AoAs and AoDs for low-overhead channel estimation. A hybrid channel estimation algorithm was proposed in \cite{wu2022hybrid} to exploit the channel sparsity properties through both the CS technique and the sparse Bayesian learning (SBL). By considering the hierarchical sparsity of the massive MIMO channel, \cite{wunder2019lowoverhead} provided a low-overhead channel estimation scheme for massive MIMO systems with analytical insights on the pilot overhead requirements. Nevertheless, owing to the time-varying complicated wireless propagation environment, the wireless channels of practical massive MIMO systems become challenging to be analytically modeled. These challenges thus hinder the application of the above traditional channel estimation methods, especially when large antenna arrays are deployed.

Recently, deep learning has been applied to wireless communication systems for performance improvement\cite{chen2022highaccuracy, xue2022beam,he2020modeldriven}. For instance, various DNNs, e.g., the fully-connected deep neural network (FC-DNN) \cite{ye2018power}, and the convolutional neural networks (CNN) \cite{dong2019deep,soltani2019deep,honkala2021deeprx}, were proposed to learn the channel characteristics based on a large amount of training data. Additionally, their variations, e.g., the complex denoising convolutional neural networks (CDnCNNs) \cite{zeng2020higher}, the convolutional blind denoising network (CBDNet) \cite{jin2020channel}, the CNN-based deep residual network (CDRN) \cite{liu2022deep}, etc., were also proposed to improve the accuracy of channel estimation. The successes of deep learning-based channel estimation methods show that DNNs are able to learn complicated channel characteristics in practical systems. However, with the continuously increasing scale of antenna arrays, existing deep learning-based methods encounter new challenges. First, to guarantee systems' spectral efficiency under limited time-frequency resources, DNN architectures are needed to be designed with superior learning capability from only a few pilots. This capability, therefore, necessitates employing more complicated DNN structures to exploit channel characteristics. Second, enlarged antenna scale and bandwidth further increase the computational complexity of existing DNN architectures.
In the prior deep learning-based methods, massive MIMO channels are usually represented in spatial-frequency domain representations due to the employment of massive antennas and wide bandwidth \cite{liu2020overcoming,dong2019deep,jiang2021dual}. Hence, the extraction of the underlying channel features from such multiple domains is essential for achieving accurate channel estimates. To deal with it, existing deep learning-based methods usually extract the multi-domain features via one DNN module. For example, \cite{soltani2019deep,dong2019deep,jiang2021dual} proposed to extract the spatial-frequency features with one CNN module. Nonetheless, different features in various domains, in fact, correspond to different correlations in various domains, which are generally irrelevant to each other. One DNN module is not able to extract multi-domain features effectively. Thereby, an advanced multi-domain feature extractor composed of multiple DNN modules for jointly extracting multi-domain features is highly desired.

In addition, it is generally true that more pilots can obtain more channel knowledge and better channel estimation performance. Therefore, the knowledge learned from the high-density pilots (e.g., obtained by accumulating low-density pilots during the same coherent interval or allocating specific slots for high-density pilot training only in the training dataset collection period) will be beneficial to improving the estimation performance of the low-overhead channel estimation if the learned knowledge can be transferred.
Fortunately, transfer learning, a.k.a. knowledge transfer, is known to be able to transfer knowledge. Specifically, transfer learning has been successfully applied to improve channel estimation accuracy under time-varying environments to overcome fast channel fluctuations \cite{yang2020deep,alves2021deep}. This consequently promises the potential of applying transfer learning to low-overhead channel estimation for performance enhancement.

Motivated by this, we first propose a dual-attention-based channel estimation network (DACEN) to realize low-overhead channel estimation by jointly learning the multi-domain features of massive MIMO channels and learning the domain-specific features via the specific attention modules. As the number of delayed paths is usually small compared to that of subcarriers, we adopt the spatial-temporal domain channel representation in DACEN to reduce its memory usage and model complexity. Particularly, the DACEN consists of two attention modules, i.e., the temporal attention module (TAM) and the spatial attention module (SAM). The TAM is designed to learn the temporal correlations among different delayed paths, while the SAM is designed to learn the correlation of CSI among different antennas. Leveraging the SAM to distinguish the highly correlated antennas and thus primarily focus on the limited number of effective antennas maintains the estimation accuracy and significantly reduces the computational complexity.
Based on the DACEN, we also propose a parameter-instance transfer learning approach to further improve the estimation accuracy by transferring the channel knowledge learned from the high-density pilots during the model training phase to that with low-density pilots.

The major contributions of this paper are summarized as follows:

\begin{enumerate}
    \item We propose a dual-attention-based channel estimation network to realize accurate channel estimation via low-density pilots for massive MIMO systems. By jointly learning the spatial-temporal domain features of massive MIMO channels with the TAM and the SAM, the proposed DACEN can achieve accurate channel estimates via low-density pilots. Besides, by exploiting the spatial correlation of massive MIMO channels, the SAM has much lower time and space complexity than the widely adopted convolutional neural networks.
    The effectiveness of the TAM also verifies that the wireless channels of massive MIMO systems may not satisfy the wide-sense stationary uncorrelated scattering (WSSUS) assumption, and the multiple paths with different delays are correlated with each other, which is beneficial to the low-overhead channel estimation.

    \item We propose a parameter-instance transfer learning approach based on the DACEN to further improve the channel estimation accuracy by transferring the channel knowledge learned from the high-density pilots during the model training phase. Note that the high-density pilots can be obtained by accumulating the low-density pilots during a coherent interval or allocating specific slots for high-density pilot training only in the training dataset collection period. By pre-training the DACEN with the obtained high-density pilots (parameter transfer) and incorporating the reweighting of low-density pilots when training the DACEN (instance transfer), CSI can be well estimated with merely low-density pilots.

    \item The transmitted pilot signal follows the 3rd generation partnership project (3GPP) 5G technical specification \cite{3gpp2023ts38211} on CSI-RS configurations for acquiring downlink channel estimates. In addition, since the CSI-RS is broadcast by the base station to all users in the system, the proposed methods can be directly applied to multi-user scenarios. The effectiveness and the superiority of the proposed methods are verified by extensive numerical experiments under various low-density pilot settings.
    
\end{enumerate}

The remainder of this paper is organized as follows. In Section \ref{Sec. Sys.}, we introduce the system model and formulate the channel estimation problem. In Section \ref{Sec. NN}, the novel DACEN architecture is proposed. To further improve the channel estimation accuracy, we propose the parameter-instance transfer learning approach in Section \ref{Sec. TL}. In Section \ref{Sec. Exp.}, the experiment results and computational complexity comparison are presented. Finally, we conclude this work in Section \ref{Sec. Con.}.

\textit{Notation}: Underlined bold uppercase letters, bold uppercase letters, and bold lowercase letters represent tensors, matrices, and vectors, respectively. $\underline{\mathbf{A}}^{(i)}$ is the $i$-th slice of the tensor $\underline{\mathbf{A}}$, $\mathbf{A}^{(i)}$ is the $i$-th row of the matrix $\mathbf{A}$, $\mathbf{A}^{(i,j)}$ represents the element of the matrix $\mathbf{A}$ at the $i$-th row and the $j$-th column, and $\mathbf{a}^{(i)}$ is the $i$-th element of the vector $\mathbf{a}$. $\operatorname{Im}\{\underline{\mathbf{A}}\}$ and $\operatorname{Re}\{\underline{\mathbf{A}}\}$ denote the imaginary part and real part of $\underline{\mathbf{A}} \in \mathbb{C}$. $\underline{\hat{\mathbf{A}}}$ represents the estimate of $\underline{\mathbf{A}}$. $\mathrm{E}(\cdot)$, $\operatorname{Var}(\cdot)$, and $\|\cdot\|_2$ denote the expectation, variance and L2 norm respectively. $\delta(\cdot)$ denotes the Dirac delta function and $\log (\cdot)$ denotes base 10 logarithm operation. $\otimes$, $<\cdot, \cdot>$, $[\cdot, ..., \cdot]$ and $\mathrm{R}(\cdot)$ denote the Hadamard product, the inner product, the concatenation operation, and the rearrange operation, respectively. To further improve the readability of this paper, the descriptions associated with some key symbols are listed in Table \ref{symbols}.

\begin{table*}[htbp]
\centering
\caption{List of key symbols and descriptions.}
\label{symbols}
\resizebox{\textwidth}{!}{%
\begin{tabular}{c|c|c|c}
\hline
\textbf{Symbol}                                                         & \textbf{Description}              & \textbf{Symbol}                                                                       & \textbf{Description}                                                                                                                         \\ \hline
$N_T$                                                                   & the number of transmitting antennas & $N_{RB}$                                                                & the total number of RBs of the system \\ \hline
$N_R$                                                                   & the number of receiving antennas    & $\rho$  & pilot density \\  \hline
$N_c$                                                                   & the number of subcarriers           & $\underline{\mathbf{Y}}_{pH}     \in \mathbb{C}^{N_R \times N_T \times N_H}$ & high-density received pilot signal (tensor form)  \\   \hline
$N_P$                                                                   & the number of considered entries to reserve after frequency-time conversion  & $\underline{\mathbf{Y}}_{pL}     \in \mathbb{C}^{N_R \times N_T \times N_L}$ & low-density received pilot signal (tensor form)   \\  \hline
$\underline{\mathbf{H}}_f \in   \mathbb{C}^{N_R \times N_T \times N_c}$ & spatial-frequency domain channel (tensor form) & $s^{th}$                                                                & cosine similarity threshold     \\ \hline
$\underline{\mathbf{H}}_t \in   \mathbb{C}^{N_R \times N_T \times N_P}$ & spatial-temporal domain channel (tensor form)  & $\mathbf{w} \in \mathbb{R}^{N_{D_{L_E}}^{train} \times 1}$  & instance weights  for   all training samples                                                                         \\ \hline
\end{tabular}
}
\end{table*}

\section{System Model and Problem Formulation}\label{Sec. Sys.}

\subsection{System Model}

In this paper, the downlink of a massive MIMO system with $N_T \gg 1$ antennas at the BS and $N_R$ antennas at the user end (UE) is considered.\footnote{Note that,  to clearly demonstrate the proposed channel estimation method, we take one of the multi-antenna UEs in the system as an example. The proposed method can be directly applied to single-cell multi-user scenarios.} The orthogonal frequency division multiplexing (OFDM) modulation with $N_c$ subcarriers is adopted. The received signal at the $i$-th subcarrier, i.e., $\mathbf{Y}^i \in \mathbb{C}^{N_R \times N_T}$, can be expressed as:
\begin{equation}\label{receiveYi}
    \mathbf{Y}^i = \mathbf{H}_f^i \mathbf{X}^i + \mathbf{N}^i,i=1,\ldots,N_c,
\end{equation}
where $\mathbf{H}_f^i \in \mathbb{C}^{N_R \times N_T}$, $\mathbf{X}^i \in \mathbb{C}^{N_T \times N_T}$, and $\mathbf{N}^i \in \mathbb{C}^{N_R \times N_T}$ denote the spatial-frequency domain channel, the diagonal matrix constituted by the transmitted signal, and the noise at the $i$-th subcarrier, respectively.

By concatenating the spatial-frequency domain channel $\mathbf{H}_f^i$ of all $N_c$ subcarriers, the obtained whole spatial frequency domain channel $\mathbf{H}_f \in \mathbb{C}^{N_R \times (N_T \times N_c)}$ can be given by:
\begin{equation}
    \mathbf{H}_f = [\mathbf{H}_f^1, ..., \mathbf{H}_f^{N_c}],
\end{equation}
which can be rewritten in a tensor form as $\underline{\mathbf{H}}_f \in \mathbb{C}^{N_R \times N_T \times N_c}$.

\subsection{Problem Formulation}

Suppose the BS transmits CSI-RS symbols (i.e., pilot symbols) and the corresponding received pilot signal at the UE is $\underline{\mathbf{Y}}_{rp} \in \mathbb{C}^{N_R \times N_T \times N_{rp}}$, $N_{rp}$ is the number of received pilots in the frequency domain, then the downlink spatial-frequency domain channel can be estimated using a DNN-based channel estimator $F_f$:
\begin{align} \label{P1}
\underline{\hat{\mathbf{H}}}_f^* &= \arg \min _{\underline{\hat{\mathbf{H}}}_f}\mathcal{L}\left(\underline{\mathbf{H}}_f - \underline{\hat{\mathbf{H}}}_f\right), \nonumber \\
\operatorname{s.t.} \ \ [\operatorname{Im}\{\underline{\hat{\mathbf{H}}}_f\},\operatorname{Re}\{\underline{\hat{\mathbf{H}}}_f\}] &= F_f([\operatorname{Im}\{\underline{\mathbf{Y}}_{rp}\},\operatorname{Re}\{\underline{\mathbf{Y}}_{rp}\}]),
\end{align}
where $\operatorname{Im}\{\cdot\}$ and $\operatorname{Re}\{\cdot\}$ denote the imaginary and real parts of the input tensor, respectively. $\mathcal{L}$ is a loss function, e.g., the mean squared error (MSE) loss function, that computes the distance between the estimated output and the expected output of a channel estimator.

Note that $N_c$ is generally large, e.g., $N_c=624$, which inevitably leads to a huge memory usage and model complexity and, thereby, an inefficient optimization procedure of (\ref{P1}). To circumvent it, a possible way is to select only a small portion of elements from $\underline{\mathbf{H}}_f$ to form a concise spatial-frequency domain channel representation, and then train the network based on this obtained concise channel representation. However, such an operation results in considerable information loss and thus degrades channel estimation performance.

\begin{figure}[htbp]
\centering
\includegraphics[width=0.48\textwidth]{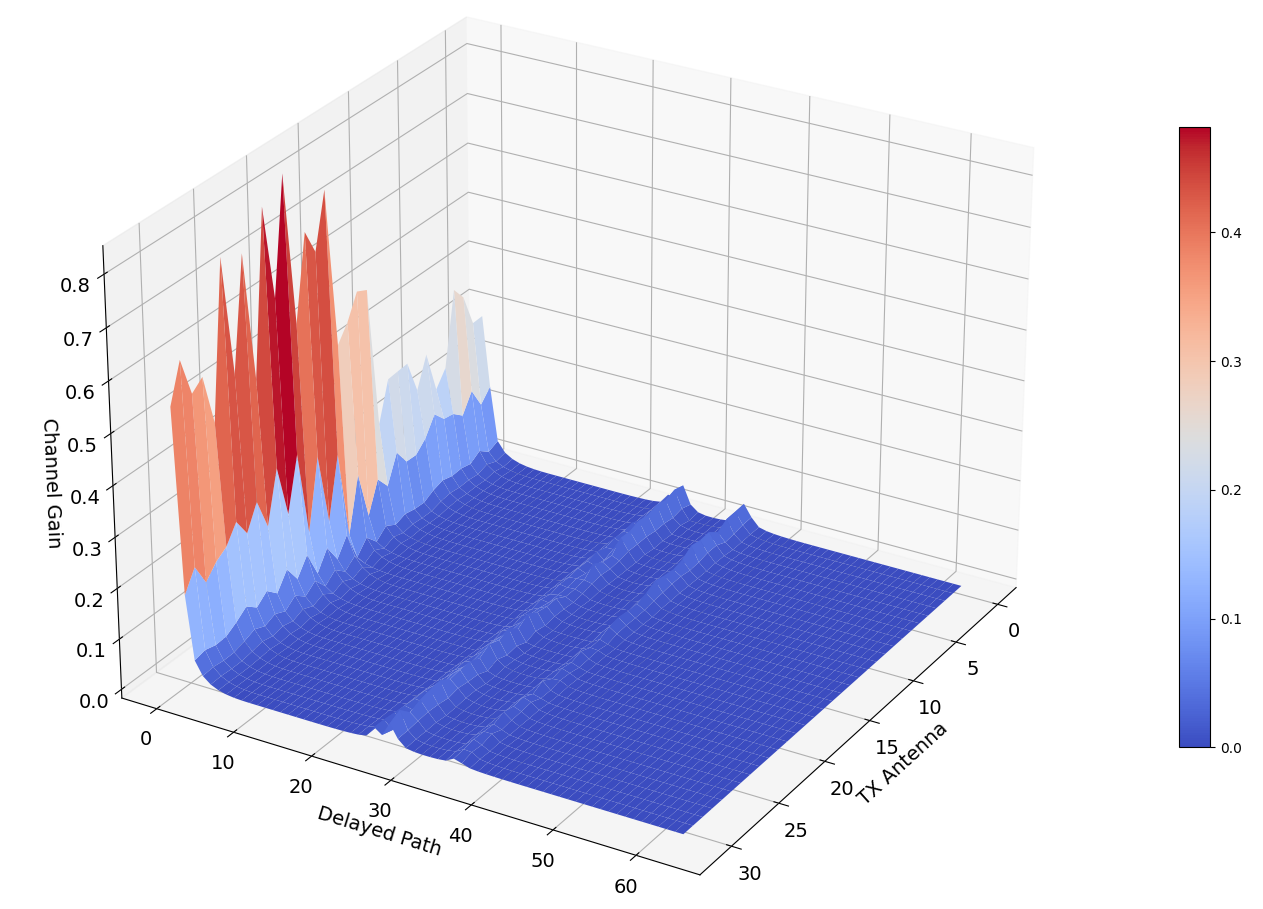}
\caption{An illustration of a spatial-temporal channel representation.}
\label{channel}
\end{figure}

Considering that the number of dominant delayed paths is usually small when compared to the number of subcarriers $N_c$ (as shown in Fig. \ref{channel},\footnote{The detailed system parameters of this spatial-temporal channel representation are provided in Section \ref{Sec. Exp.}.} there are less than $10$ dominant delayed paths in the exampled channel snapshot while we have $N_c=624$ in the frequency domain corresponding to the OFDM modulation), we first propose to deal with the hefty memory and model complexity resulting from large $N_c$ by resorting to the spatial-temporal domain channel $\mathbf{H}_t \in \mathbb{C}^{(N_R \times N_T) \times N_P}$ via a frequency-time conversion:
\begin{equation} \label{IFFT}
    \mathbf{H}_t = \text{IFFT}_{N_\text{ifft}}(\mathbf{H}_f^\prime)\mathbf{A},
\end{equation}
where $\text{IFFT}_{N_\text{ifft}}(\cdot)$ denotes the $N_\text{ifft}$-point Inverse Fast Fourier Transform (IFFT) along the frequency dimension,\footnote{In comparison with the huge memory usage and model complexity brought by a large number of subcarriers, the computational complexity of the additional IFFT operation is negligible.} and $\mathbf{H}_f^\prime \in \mathbb{C}^{(N_R \times N_T) \times N_c}$ is reshaped from $\underline{\mathbf{H}}_f \in \mathbb{C}^{N_R \times N_T \times N_c}$. $\mathbf{A} \in \mathbb{C}^{N_\text{ifft} \times N_P}$ is a right truncating matrix defined as:
\begin{equation}
\mathbf{A}=\left[\begin{array}{c}
\mathbf{I}_{N_P} \\
\mathbf{0}_{N_\text{ifft} - N_P, N_P}
\end{array}\right]
\end{equation}
where $\mathbf{I}$ is the identity matrix, $\mathbf{0}$ is the zero matrix, and $N_P$ is the number of considered entries of $\text{IFFT}_{N_\text{ifft}}(\mathbf{H}_f^\prime)$ to reserve. $\mathbf{H}_t \in \mathbb{C}^{(N_R \times N_T) \times N_P}$ can be rewritten in its tensor form as $\underline{\mathbf{H}}_t \in \mathbb{C}^{N_R \times N_T \times N_P}$. Thanks to the path loss and the limited number of scatterers in the propagation environment, only a few propagation paths within a limited delay spread can be observed at the receiving end after the transmitted signal traverses the propagation environment. Denote the channel sampling frequency as $f_s$, we assume that the maximum path delay is $N_P / f_s$, indicating that the entries behind the first $N_P$ entries may have very small path gains. Therefore, a sufficiently large $N_P$ in (\ref{IFFT}) will reserve the majority, or even the whole information, of the channel.\footnote{$N_P$ can be determined by a coarse estimate of the delay spread in the propagation environment based on the pre-collected data. As the model is deployed at the UEs located in a relatively static communication environment, the maximum delay of propagation paths will not change significantly. Therefore, $N_P$ can also be adopted in the test period.} It is also worth noting that due to the limited delay spread, $N_P$ can still be small enough compared to $N_c$, thereby effectively reducing the memory usage and model complexity.

Then, the downlink spatial-frequency domain channel estimation problem can be transformed into the downlink spatial-temporal domain channel estimation problem, i.e.,
\begin{align} \label{P2}
\underline{\hat{\mathbf{H}}}_t^* &= \arg \min _{\underline{\hat{\mathbf{H}}}_t}\mathcal{L}\left(\underline{\mathbf{H}}_t - \underline{\hat{\mathbf{H}}}_t\right), \nonumber \\
\operatorname{s.t.} \ \ [\operatorname{Im}\{\underline{\hat{\mathbf{H}}}_t\},\operatorname{Re}\{\underline{\hat{\mathbf{H}}}_t\}] &= F_t([\operatorname{Im}\{\underline{\mathbf{Y}}_{rp}\},\operatorname{Re}\{\underline{\mathbf{Y}}_{rp}\}]),
\end{align}
where $F_t$ is a DNN-based channel estimator. Compared with $\underline{\mathbf{H}}_f$, using $\underline{\mathbf{H}}_t$ as the label to train the DNN-based channel estimators can reduce tremendous memory usage and model parameters and avoid significant information loss of the propagation channel. It is worth mentioning that $F_t$ takes the received pilot signal (spatial-frequency domain representation) as the input and outputs the estimated spatial-temporal domain channel. Therefore, $F_t$ is capable of converting the signal in the spatial-frequency domain to the channel in the spatial-temporal domain implicitly.

Furthermore, as we mentioned before, besides the significantly increased complexity of existing DNN architectures, the pilot training overhead of massive MIMO systems also increases rapidly with the continuously increasing antenna array scale, especially for the downlink under frequency duplex mode (FDD). In general, the pilot training overhead of the downlink channel estimation in FDD massive MIMO systems increases linearly with the number of BS's antennas, which is, in fact, unaffordable in practical scenarios. Although there are some existing techniques that can be leveraged to reduce the pilot training overhead\cite{noh2014pilot,gao2016structured}, investigating low-overhead deep learning-based channel estimation methods is essential to improve channel estimation accuracy and reduce the pilot training overhead simultaneously for massive MIMO systems.

Note that after estimating the downlink channel, the UE generally needs to pre-process the CSI first, then compress and quantize it, and finally feed the compressed CSI back to the BS, such that the BS can conduct sophisticated signal processing or downlink precoding. In this paper, we mainly focus on accurate channel estimation with low-density pilots before the CSI feedback.\footnote{Existing CSI feedback schemes usually assume perfect downlink CSI\cite{wen2018deep,guo2020convolutional,guo2022overview} or rely on the eigenvector of the estimated downlink channel\cite{3gpp2020ts38214,liu2021evcsinet,chen2022deep}. Therefore, achieving accurate downlink channel estimates first is essential before performing the CSI feedback.}

\subsection{Pilot Density}
In 5G systems, the CSI-RS symbols are configured on resource blocks (RBs) to support channel estimation\cite{3gpp2023ts38211}. Since the CSI-RS symbols are only configured on a certain proportion of RBs, we define the pilot density $\rho$ as:
\begin{equation} \label{density}
    \rho = N_{\scaleto{CSI-RS}{5pt}} / N_{RB},
\end{equation}
where $N_{\scaleto{CSI-RS}{5pt}}$ is the number of RBs configured with CSI-RS symbols and $N_{RB}$ is the total number of RBs of the system. For clear statements in the following sections, we use two CSI-RS configurations corresponding to two pilot density aliases:
\begin{enumerate}
\item \textbf{High-density pilots with $\rho_H=N_H / N_{RB}$}: CSI-RS symbols are configured on $N_H$ RBs. The high-density pilots are used for performance comparison and for transfer learning and will be introduced in Section \ref{Sec. TL}. Note that as defined in \cite{3gpp2023ts38211}, the pilot density is generally set as $1$ or $0.5$. Therefore, we set $\rho_H=0.5$ (i.e., $N_H = 0.5 N_{RB}$) as the upper bound of the pilot density for comparison.

\item \textbf{Low-density pilots with $\rho_L=N_L / N_{RB}$}: CSI-RS symbols are configured on $N_L$ RBs, and $N_L < N_H$.
\end{enumerate}

We denote the received pilot signals under the above two configurations as $\underline{\mathbf{Y}}_{pH} \in \mathbb{C}^{N_R \times N_T \times N_H}$ and $\underline{\mathbf{Y}}_{pL} \in \mathbb{C}^{N_R \times N_T \times N_L}$, respectively.

With the objective of low-overhead channel estimation, the DNNs should be designed to conduct channel estimation with low-density CSI-RS configurations. Coupled with the enlarged antenna scale and bandwidth, a significant increase in the complexity of the estimator is generally required to attain accurate channel estimates. As such, the design of an effective network architecture with low time and space complexity for low-overhead channel estimation in massive MIMO systems is of paramount importance.

From the exampled spatial-temporal channel representation shown in Fig. \ref{channel}, both amplitude correlations between adjacent delayed paths and spatial correlations exist in massive MIMO channels. For example, the channel gains of adjacent delayed paths near the index $10$ are highly related. Based on the channel gain's fluctuations along the axis labeled `TX Antenna', the CSI correlations in the spatial domain can also be clearly observed. Fig. \ref{channel}  also indicates that the temporal domain features (i.e., the correlations among delayed paths) and the spatial domain features (i.e., the correlations among the CSI at different antennas) are irrelevant to each other.
As such, it is anticipated that the performance of jointly extracting different domain features via an advanced multi-domain feature extractor consisting of multiple dedicatedly designed DNN modules will outperform extracting the multi-domain features via one DNN module since different domain feature extractions generally necessitate different network architecture designs to match their specific characteristics. In the following section, we propose the DACEN to learn multi-domain features jointly with SAMs and subsequent TAMs.

\section{Dual-Attention Channel Estimation Network} \label{Sec. NN}
 To fully exploit the characteristics of the spatial-temporal domain channel, the DACEN is proposed in this section by exclusively learning the time domain and the spatial domain features with the TAM and the SAM, respectively.
\subsection{Overall Network Architecture}

\begin{figure*}[htbp]
\centering
\includegraphics[width=0.9\textwidth]{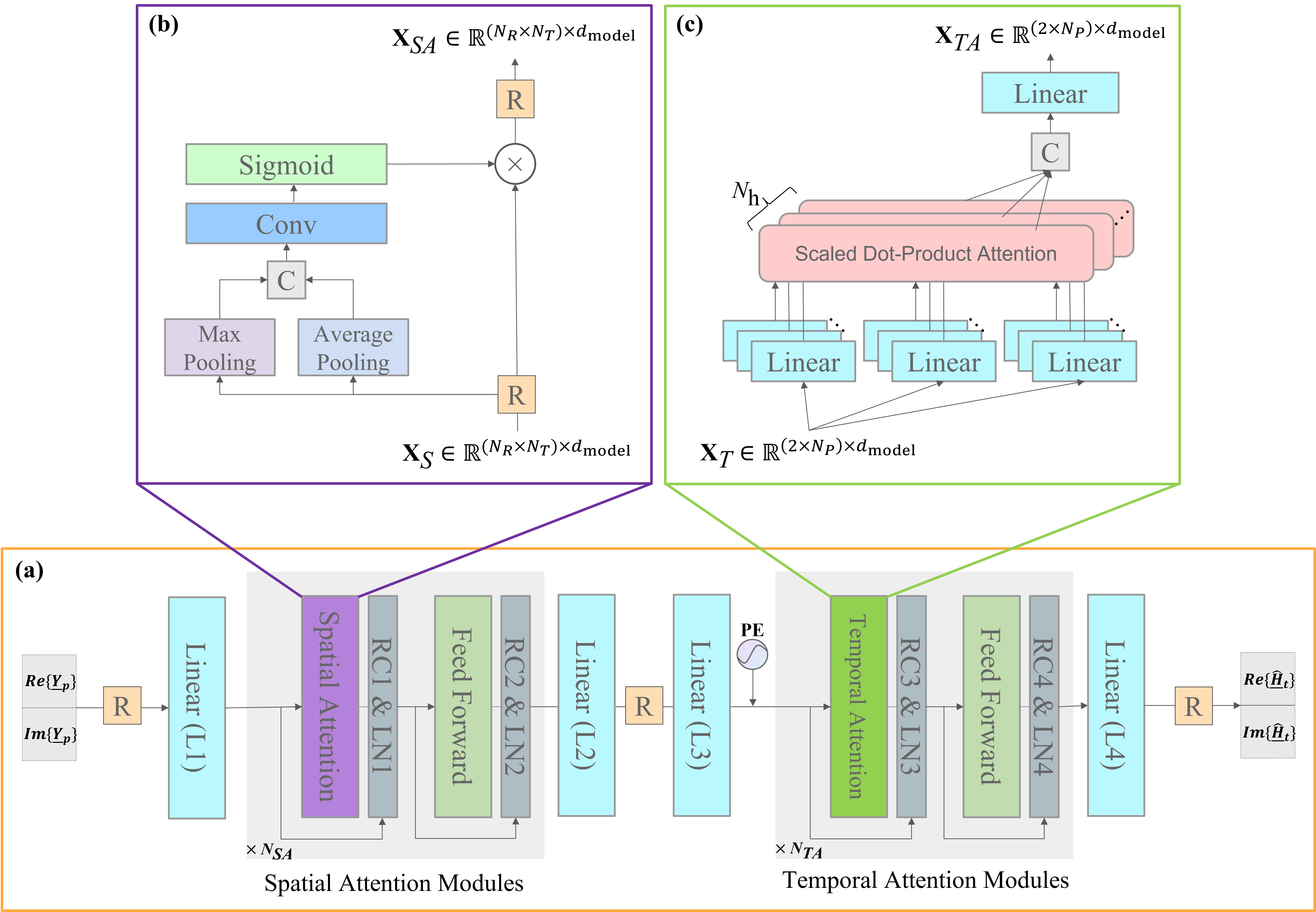}
\caption{Proposed dual-attention channel estimation network (DACEN). (a). The overall network architecture of the DACEN. (b). The spatial attention mechanism. (c). The temporal attention mechanism. \textbf{RC}: residual connection. \textbf{LN}: layer normalization. \textbf{PE}: positional encoding. \textbf{C}: concatenation operation. \textbf{R}: rearrangement operation.}
\label{nn}
\end{figure*}

As shown in Fig. \ref{nn}, the proposed DACEN is mainly composed of $N_{SA}$ SAMs and $N_{TA}$ TAMs, and each attention module consists of one attention layer, one feed-forward layer, as well as additional residual connection and layer normalization operations. The TAM is utilized to learn temporal dependencies among different delayed paths, while the SAM is designed to extract spatial domain features. Before the first SAM, the first TAM, and the output, there are linear layers (i.e., `L1', `L2', `L3', and `L4' in Fig. \ref{nn} (a)) and rearrangement operations (represented by a box labeled `R') to map the input to appropriate tensor dimensions. Particularly, `L1' and `L3' map the original tensors to high-dimensional feature spaces, enabling better feature learning. The outputs of `L1' and `L3' are denoted as $\mathbf{X}_{L1} \in \mathbb{R}^{(N_R \times N_T) \times d_\text{model}}$ and $\mathbf{X}_{L3} \in \mathbb{R}^{(2 \times N_P) \times d_\text{model}}$, respectively. In addition, the sinusoidal positional encoding $\mathbf{P} \in \mathbb{R}^{(2 \times N_P) \times d_\text{model}}$ with the same shape as $\mathbf{X}_{L3}$ is inserted before the TAMs to facilitate their learning of the positional information of delayed paths. Below we introduce the basic components of the proposed DACEN in detail.

\subsection{Temporal Attention Module}

The temporal attention layer in TAM is designed by exploiting the advantages of the multi-head attention \cite{vaswani2017attention} proposed for extracting long-term dependencies in sentences.
The multi-head attention has been extended to other sequential data such as time series \cite{zhou2022interpretable} and image patches \cite{dosovitskiy2021image}. As the same transmitted pilot signals traverse across multiple scatterers and result in multiple paths, we propose to leverage the multi-head attention to learn the temporal dependencies from these multiple paths to facilitate accurate channel estimation. Denote the input of the temporal attention layer as $\mathbf{X}_T \in \mathbb{R}^{(2 \times N_P) \times d_\text{model}}$, where the real and imaginary part of the original complex representation are concatenated together, and $d_\text{model}$ denotes the input representation dimension.\footnote{The representation dimension is empirically determined through hyper-parameter tuning.} The operations of the temporal attention layer are formulated as (\ref{TA_1}) - (\ref{TA_5}). In particular, the input is first projected to $N_h$ (the number of attention heads) sets of query, key, and value matrices for calculating head-wise attention. The output of each attention head, denoted by $\mathbf{O}_{i} \in \mathbb{R}^{(2 \times N_P) \times d_v}$, can be calculated as:
\begin{equation} \label{TA_1}
\mathbf{O}_{i} = \text{Attention}\left(\mathbf{Q}_i, \mathbf{K}_i, \mathbf{V}_i\right), i=1, 2, ... , N_h,
\end{equation}
$\text{Attention}(\cdot)$ denotes the scaled dot-product attention taking the query matrix $\mathbf{Q}_i \in \mathbb{R}^{(2 \times N_P) \times d_k}$, key matrix $\mathbf{K}_i \in \mathbb{R}^{(2 \times N_P) \times d_k}$ and value matrix $\mathbf{V}_i \in \mathbb{R}^{(2 \times N_P) \times d_v}$ as inputs. By mapping from the high-dimensional feature representation of the spatial-temporal domain channels, we obtain $\mathbf{Q}_i$, $\mathbf{K}_i$, and $\mathbf{V}_i$ as:
\begin{align} \label{projection}
    \mathbf{Q}_i &= \mathbf{X}_T \mathbf{W}_{i}^{Q}, \\
    \mathbf{K}_i &= \mathbf{X}_T \mathbf{W}_{i}^{K}, \\
    \mathbf{V}_i &= \mathbf{X}_T \mathbf{W}_{i}^{V},
\end{align}
where $\mathbf{W}_{i}^{Q} \in \mathbb{R}^{d_\text{model} \times d_k}$, $\mathbf{W}_{i}^{K} \in \mathbb{R}^{d_\text{model} \times d_k}$, and $\mathbf{W}_{i}^{V} \in \mathbb{R}^{d_\text{model} \times d_v}$ are learnable weights, and $d_k = d_v = d_\text{model}/N_h$. The scaled dot-product attention $\text{Attention}(\cdot)$ is defined as:
\begin{equation}\label{scaled}
\text{Attention}(\mathbf{Q}_i, \mathbf{K}_i, \mathbf{V}_i)=\operatorname{softmax}\left(\frac{\mathbf{Q}_i \mathbf{K}_i^{T}}{\sqrt{d_{k}}}\right) \mathbf{V}_i,
\end{equation}
where $\operatorname{softmax}(\cdot)$ is an activation function. Finally, the outputs of all attention heads are concatenated (represented by a box labeled `C') to form the output $\mathbf{X}_{TA} \in \mathbb{R}^{(2 \times N_P) \times d_\text{model}}$ of the temporal attention layer:
\begin{equation}  \label{TA_5}
    \mathbf{X}_{TA} = [\mathbf{O}_{1}, \ldots, \mathbf{O}_{N_h}]\mathbf{W}^O,
\end{equation}
where $\mathbf{W}^{O} \in \mathbb{R}^{d_\text{model} \times d_\text{model}}$ is a learnable weight matrix. It is worth mentioning that by projecting the original representation of delayed paths into different representation sub-spaces as in (\ref{projection}), the temporal attention layer is expected to learn richer features than simply learning from the original representation space since, intuitively, we can obtain more information about an object when we observe it from different views simultaneously.

Normalization techniques are effective at stabilizing and reducing the training time of neural networks\cite{ba2016layer}, and we utilize layer normalization in the DACEN  for this purpose. Residual connection, proposed initially in \cite{he2016deep} to avoid gradient vanishing/exploding, is also adopted in DACEN to enable deep neural network connections. Specifically, two combos of layer normalization and residual connection are applied after the attention layer and the feed-forward layer, respectively. Among these, the feed-forward layer, composed of two linear layers and one ReLU activation in between, is connected right after the first residual connection. Denoting the outputs of the first layer normalization, the first residual connection, the feed-forward layer, the second layer normalization, and the second residual connection as $\mathbf{X}_{LN1} \in \mathbb{R}^{(2 \times N_P) \times d_\text{model}}$, $\mathbf{X}_{RC1} \in \mathbb{R}^{(2 \times N_P) \times d_\text{model}}$, $\mathbf{X}_{FF} \in \mathbb{R}^{(2 \times N_P) \times d_\text{model}}$, $\mathbf{X}_{LN2} \in \mathbb{R}^{(2 \times N_P) \times d_\text{model}}$ and $\mathbf{X}_{RC2} \in \mathbb{R}^{(2 \times N_P) \times d_\text{model}}$, respectively, we have:
\begin{equation}
     \mathbf{X}_{RC1} = \mathbf{X}_{TA} + \mathbf{X}_T,
\end{equation}
\begin{equation} \label{ln1}
\mathbf{X}_{LN1} = \frac{\mathbf{X}_{RC1}-\mathrm{E}_{-1}[\mathbf{X}_{RC1}]}{\sqrt{\operatorname{Var}_{-1}[\mathbf{X}_{RC1}]+\epsilon}} \otimes \mathbf{g}_{LN1} + \mathbf{b}_{LN1},
\end{equation}
where $\mathrm{E}_{-1}(\cdot)$ and $\operatorname{Var}_{-1}(\cdot)$ represent taking the expectation and variance of the input matrix along its last dimension, and $\otimes$ denotes the Hadamard product. $\mathbf{g}_{LN1} \in \mathbb{R}^{1 \times d_\text{model}}$ and $\mathbf{b}_{LN1} \in \mathbb{R}^{1 \times d_\text{model}}$ are learnable affine transformation parameters, and $\epsilon$ is a small number (e.g. 1e-5) used for numerical stability.

The output of the feed-forward layer is given by
\begin{equation}\label{FF}
     \mathbf{X}_{FF}=\left(\operatorname{ReLU}(\mathbf{X}_{RC1}\mathbf{W}_1+\mathbf{b}_1)\right)\mathbf{W}_2+\mathbf{b}_2,
\end{equation}
where $\operatorname{ReLU}(\cdot)$ is an activation function, $\mathbf{W}_1 \in \mathbb{R}^{d_\text{model} \times d_\text{ff}}$ and $\mathbf{W}_2 \in \mathbb{R}^{d_\text{ff} \times d_\text{model}}$ are the learnable weights and $\mathbf{b}_1 \in \mathbb{R}^{1 \times d_\text{ff}}$ and $\mathbf{b}_2 \in \mathbb{R}^{1 \times d_\text{model}}$ are the learnable biases of the first and the second linear layers, respectively. $d_\text{ff}$ is the inner dimension of the feed-forward layer. Empirically, a large $d_\text{model}$ (e.g., 512) and $d_\text{ff} \ge d_\text{model}$ enable more effective feature learning. However, to restrict the computational complexity of the feed-forward layer, we set $d_\text{ff}=d_\text{model}$ in this work. Then, based on $\mathbf{X}_{FF}$, we have:

\begin{equation}
     \mathbf{X}_{RC2} = \mathbf{X}_{FF} + \mathbf{X}_{LN1},
\end{equation}
\begin{equation} \label{ln2}
\mathbf{X}_{LN2}=\frac{\mathbf{X}_{RC2}-\mathrm{E}_{-1}[\mathbf{X}_{RC2}]}{\sqrt{\operatorname{Var}_{-1}[\mathbf{X}_{RC2}]+\epsilon}} \otimes \mathbf{g}_{LN2} + \mathbf{b}_{LN2},
\end{equation}
where $\mathbf{g}_{LN2} \in \mathbb{R}^{1 \times d_\text{model}}$ and $\mathbf{b}_{LN2} \in \mathbb{R}^{1 \times d_\text{model}}$ are learnable affine transformation parameters.

In addition, according to (\ref{scaled}), the scaled dot-product attention attends to different delayed paths equivalently, which disregards their positional information. To exploit the positional information of delayed paths, we add a sinusoidal positional encoding to the output of the linear layer `L3', i.e., $\mathbf{X}_{L3} \in \mathbb{R}^{(2 \times N_P) \times d_\text{model}}$, before it is input to the first TAM, as shown in Fig \ref{nn}. Note that the sinusoidal positional encoding has the same shape as $\mathbf{X}_{L3}$ to provide unique positional information for delayed paths at different positions. Specifically, the sinusoidal positional encoding $\mathbf{P} \in \mathbb{R}^{(2 \times N_P) \times d_\text{model}}$ is generated by:
\begin{align}
    \mathbf{P}^{(p, 2i)} &= \sin\left(\frac{p}{\omega^{2i/d_\text{model}}}\right),\\
    \mathbf{P}^{(p, 2i+1)} &= \cos\left(\frac{p}{\omega^{2i/d_\text{model}}}\right),
\end{align}
where $p \in [0, 2 \times N_P - 1]$ is the index along the $2 \times N_P$ dimension, $i \in [0, d_\text{model} / 2 - 1]$ indicates the index along the $d_\text{model}$ dimension, and $\omega$ is a hyper-parameter related to the number of delayed paths. After that, the input of the first TAM, denoted by $\mathbf{I}_{TA1}$, is obtained as:
\begin{equation} \label{addPE}
    \mathbf{I}_{TA1} = \mathbf{X}_{L3} + \mathbf{P},
\end{equation}
while the input of other TAM except the first TAM is the output of its previous TAM.

The computational complexity of the TAM mainly comes from the temporal attention layer and the feed-forward layer. According to (\ref{TA_1}) - (\ref{TA_5}) and (\ref{FF}),  the computational complexity of the TAM can be expressed by:

\begin{align} \label{tam_complexity}
\mathcal{O}(\text{TAM}) &= \mathcal{O}(3 (2 N_P) (d_\text{model})^2 + 2(2 N_P)^2 d_\text{model}\nonumber \\ 
& \quad\quad\quad + (2 N_P) (d_\text{model})^2 + 2 (2 N_P) d_\text{ff} d_\text{model}) \nonumber \\
 &\overset{(a)}{=} \mathcal{O}(6 (2 N_P) (d_\text{model})^2 + 2(2 N_P)^2 d_\text{model}),
\end{align}
where $(a)$ comes from  $d_\text{ff}=d_\text{model}$.

\subsection{Spatial Attention Module}

Owing to the generally collocated antenna arrays and the limited angular spread of the incident signals, massive MIMO channels always exhibit spatial correlations. In addition, different antenna separations lead to different levels of spatial correlation \cite{liu2020overcoming}. Therefore, it is possible to focus mainly on the highly correlated antennas while achieving good estimation accuracy but greatly reducing the computational complexity of the channel estimation. 
Although such spatial correlation features can be extracted by CNNs, the computational complexity of the CNN layer rapidly increases with the increasing antenna scale and may become unbearable in practical communication systems. To this end, we resort to a spatial attention mechanism \cite{woo2018cbam} and propose a powerful and efficient spatial domain feature extractor, i.e., the SAM.

The spatial attention layer in SAM is specified as follows. The input $\mathbf{X}_S \in \mathbb{R}^{(N_R \times N_T) \times d_\text{model}}$ is first rearranged to the tensor $\underline{\mathbf{X}}_S^\prime \in \mathbb{R}^{N_R \times N_T \times d_\text{model}}$. Then, both the max-pooling and the average-pooling operations are employed to generate two feature maps $\mathbf{X}_S^{max} \in \mathbb{R}^{N_R \times N_T \times 1}$ and $\mathbf{X}_S^{avg} \in \mathbb{R}^{N_R \times N_T \times 1} $ along the representation dimension from $\underline{\mathbf{X}}_S^\prime$. These two feature maps are then concatenated and processed by a convolutional layer and a sigmoid activation function to generate one spatial attention map. After that, conducting the Hadamard product between the spatial attention map and the input yields the output $\underline{\mathbf{X}}_{SA}^{\prime\prime} \in \mathbb{R}^{N_R \times N_T \times d_\text{model}}$. Finally, another rearrangement operation is applied to rearrange $\underline{\mathbf{X}}_{SA}^{\prime\prime}$ to $\mathbf{X}_{SA} \in \mathbb{R}^{(N_R \times N_T) \times d_\text{model}}$. Concretely, the operations of the spatial attention layer can be formulated as (\ref{SA_1}) to (\ref{SA_3}): 
\begin{equation}
\underline{\mathbf{X}}_S^\prime = \mathrm{R}(\mathbf{X}_{S}),\label{SA_1}
\end{equation}
\begin{equation}
\underline{\mathbf{X}}_S^{\prime\prime}=\sigma\left(\operatorname{Conv}^{1\times1}\left([\operatorname{MP}(\underline{\mathbf{X}}_S^\prime),\operatorname{AP}(\underline{\mathbf{X}}_S^\prime)]\right)\right) \otimes \underline{\mathbf{X}}_S^\prime,
\end{equation}
\begin{equation}
\mathbf{X}_{SA} = \mathrm{R}(\underline{\mathbf{X}}_{S}^{\prime\prime}),\label{SA_3}
\end{equation}
where $\sigma(\cdot)$ denotes the sigmoid activation function, $\operatorname{Conv}(\cdot)^{1\times1}$ is a convolutional layer with an $1 \times 1$ filter, and $\operatorname{MP}(\cdot)$ and $\operatorname{AP}(\cdot)$ represent the max-pooling and average-pooling, respectively.

Moreover, to further enhance the learning capability of the SAM, we put one feed-forward layer after the spatial attention mechanism inspired by the Transformer architecture\cite{vaswani2017attention}, and the residual connection and layer normalization operations are wrapped around the spatial attention mechanism and the feed-forward layer.
Thanks to the specifically designed architecture, the SAM can flexibly capture the global and local features of massive MIMO channels in the spatial domain in one step. Besides, compared with CNNs under the same conditions, the SAM has less complexity and fewer parameters.
The computational complexity of the SAM can be calculated as:
\begin{align} \label{sam_complexity}
    \mathcal{O}(\text{SAM}) &= \mathcal{O}(2 N_R N_T  + N_R N_T d_\text{model} + 2 N_R N_T d_\text{ff} d_\text{model}) \nonumber \\
    &\overset{(a)}{=} \mathcal{O}(2 N_R N_T  + N_R N_T d_\text{model} + 2 N_R N_T (d_\text{model})^2),
\end{align}
where $(a)$ comes from $d_\text{ff}=d_\text{model}$ as claimed in Section \ref{Sec. NN}.B.

As a comparison, when a standard two-dimensional CNN (i.e., the spatial convolution layer (SConv)) is applied with a $K_S \times K_S$ (e.g., $3\times3$, or $7\times7$) filter\footnote{The filter size is usually set larger than 1 to support sparse connections and keep spatial locality and finally guarantee the performance of CNNs.} and a representation dimension $d_\text{model}$ for spatial domain feature extraction, the computational complexity of the SConv is:
\begin{equation} \label{sconv_complexity}
\mathcal{O}(\text{SConv}) = \mathcal{O}(N_R N_T (K_S)^2 (d_\text{model})^2).
\end{equation}
Comparing  (\ref{sconv_complexity}) with (\ref{sam_complexity}), it can be found that the SAM is more computationally efficient than the SConv.

\subsection{Learning Strategy}
Given the low-density pilot samples and corresponding channel labels for training, i.e., $D_L^{train} = \{(\underline{\mathbf{Y}}_{pL,i},\underline{\mathbf{H}}_{t,i})|i=1,...,N_{D_L}^{train}\}$, the proposed DACEN can be trained from scratch by minimizing the mean square error (MSE) loss as (\ref{scratch_loss}) at the top of the next page shows, where $N_{D_L}^{train}$ is the number of low-density samples for training.

\begin{figure*}
\begin{align} \label{scratch_loss}
F_L^{Scr} = \arg \min _{F_L} \frac{1}{N_{D_L}^{train}} \sum_{i=1}^{N_{D_L}^{train}} \|[\operatorname{Im}\{\underline{\mathbf{H}}_{t,i}\},\operatorname{Re}\{\underline{\mathbf{H}}_{t,i}\}] - F_L([\operatorname{Im}\{\underline{\mathbf{Y}}_{pL,i}\},\operatorname{Re}\{\underline{\mathbf{Y}}_{pL,i}\}])\|_2^2,
\end{align}
\hrulefill
\end{figure*}

\begin{figure*}[htbp]
\centering
\includegraphics[width=0.9\textwidth]{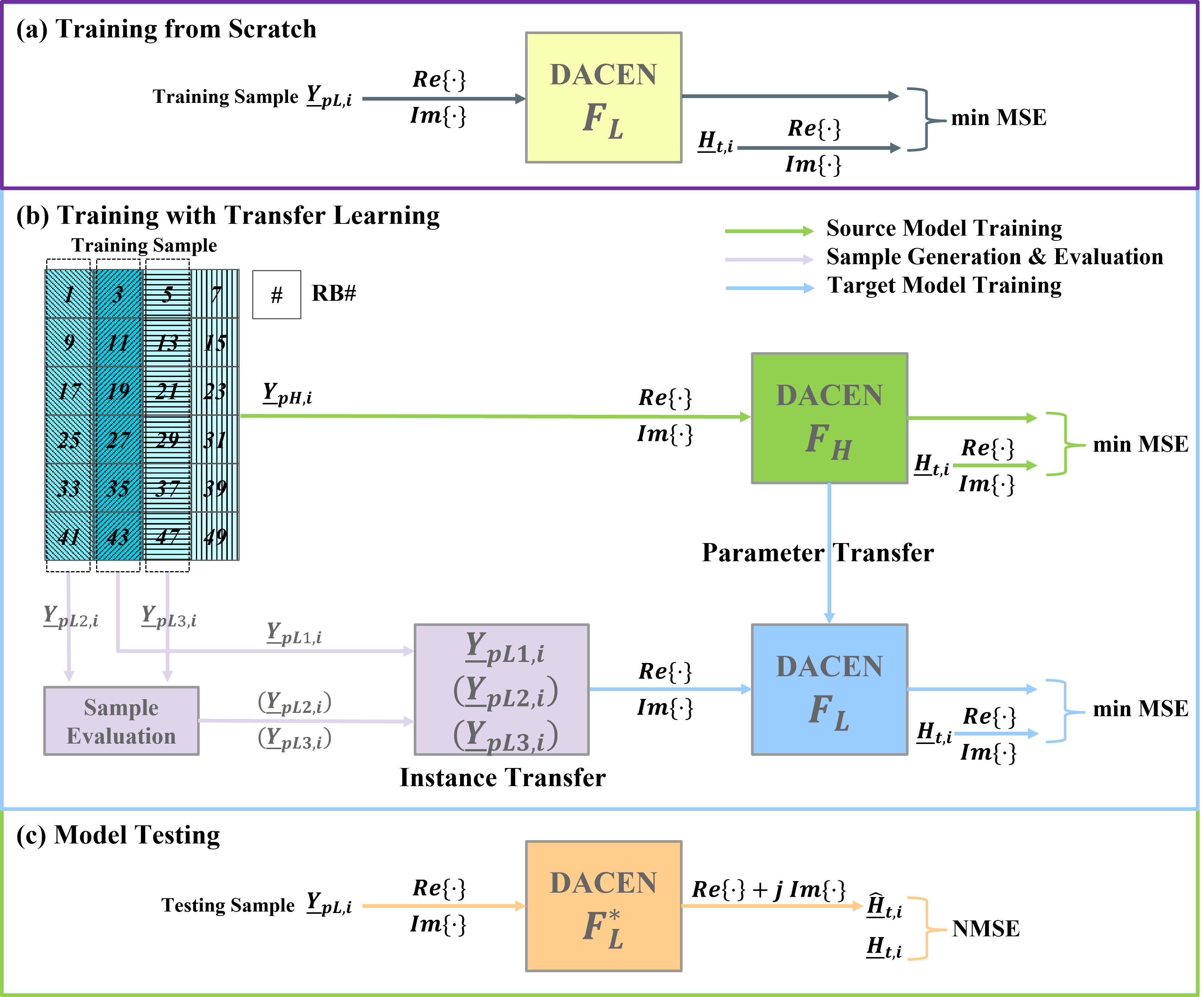}
\caption{ Training and testing processes of the DACEN. (a). Training the DACEN from scratch. (b) Training the DACEN with the parameter-instance transfer learning approach. (c). Testing the DACEN with low-density pilots.}
\label{tf}
\end{figure*}

The process of training the DACEN from scratch is shown in Fig. \ref{tf} (a). 
Considering that more pilots are generally beneficial to obtain more accurate channel estimates, based on the pre-acquired received high-density pilots $\underline{\mathbf{Y}}_{pH}  \in \mathbb{C}^{N_R \times N_T \times N_H}$ ($N_H > N_L$) (e.g., obtained by allocating specific slots for high-density pilot training during the training dataset collection period), it is promising to employ transfer learning to train the DACEN for better estimation performance. In the following section, we propose a parameter-instance transfer learning approach to enhance the estimation performance of the DACEN further.

\section{Parameter-Instance Transfer Learning for Low-overhead Channel Estimation} \label{Sec. TL}

Generally, estimating the channel with more pilots would achieve more accurate channel estimates than with fewer pilots since more knowledge of the channel can be learned from more pilots. Therefore, we exploit transfer learning to transfer the knowledge of the channel learned from high-density pilots pre-acquired during the training dataset collection period to improve further the accuracy of our proposed DACEN-based channel estimation method. In this section, we first briefly introduce the background of transfer learning and the common transfer learning approaches. After that, we introduce the proposed parameter-instance transfer learning approach.

\subsection{Transfer Learning}
The basic assumption of machine learning methodologies is that the training data and testing data are sampled from the same domain, which indicates that the two data sets are in the same feature space and have the same distribution\cite{pan2010survey,weiss2016survey}. However, the assumption may not hold in many real-world scenarios. 
As a consequence, transfer learning is designed to improve the learner for a target task by transferring knowledge from a source task to accommodate the situations when the basic assumption does not hold\cite{pan2010survey}.
Suppose a source domain $S$, a corresponding source task $\mathcal{T}_S$ and a learner $F_S$ for $\mathcal{T}_S$. A training dataset collected from the source domain $S$, denoted by $\mathcal{D}_S$, can be used for training the learner $F_S$ for the source task $\mathcal{T}_S$. When the data distribution changes, a new data domain and a corresponding task emerge, which we denote as $T$ and $\mathcal{T}_T$, respectively. Recollecting a large training dataset from the target domain $T$ might be unaffordable or impossible. Therefore, a learner $F_T$ for the target task $\mathcal{T}_T$ can only be obtained with a small portion of data or even without data from the target domain $T$. In response to such situations, transfer learning, also known as knowledge transfer, is designed to improve the learner $F_T$ for $\mathcal{T}_T$ by transferring knowledge from the source domain data $\mathcal{D}_S$.

Based on “what to transfer,” there are four typical approaches in transfer learning \cite{pan2010survey}, including the instance-transfer approach, the parameter-transfer approach, the feature representation-transfer approach, and the relational knowledge-transfer approach.

$1)$ \textbf{Instance-transfer approach}. Assuming that some data in the source domain have the potential to facilitate the target task, the instance-transfer approach aims at reusing such data for learning a learner for the target task, usually by instance reweighting and importance sampling.

$2)$ \textbf{Parameter-transfer approach}. The parameter-transfer approach assumes that source tasks and target tasks may share some parameters or prior distributions of the hyper-parameters of the learners. Therefore, the knowledge to transfer can be encoded into shared parameters or priors. In the context of deep transfer learning, the well-known pre-training method is a representative parameter-transfer approach.

$3)$ \textbf{Feature representation-transfer approach}. The feature representation-transfer approach is to directly encode the knowledge to a good feature representation for the target task. In other words, such feature representation should be domain invariant and is thus expected to improve the performance of the target task.

$4)$ \textbf{Relational knowledge-transfer approach}. The relational knowledge-transfer approach is designed for knowledge transfer across relational domains based on the assumption that some relationships among the data in relational domains are similar, where the knowledge to be transferred is intuitively the relationships among the data.

As far as we know, the parameter-transfer approach has been applied in channel prediction under time-varying environments by transferring environmental knowledge from previously experienced environments\cite{yang2020deep,alves2021deep}, while the others are rarely exploited for channel estimation yet. Since the feature representation-transfer approach requires dedicated feature selection processes and the relational knowledge-transfer approach applies only to knowledge transfer within relational domains, they can not be adopted to improve low-overhead channel estimation straightforwardly. In contrast, considering that more knowledge of the channel can be learned from more pilots, the parameter-transfer approach and the instance-transfer approach are promising to fully transfer knowledge of the channel by pre-training and augmenting more training data from the pre-acquired high-density pilots. Therefore, we propose the parameter-instance transfer learning approach to improve the channel estimation accuracy of the DACEN-based channel estimation method.

\subsection{Proposed Parameter-Instance Transfer Learning Approach}

Notably, the proposed parameter-instance transfer learning approach benefits from both parameter transfer and instance transfer. On the one hand, the parameters of a DACEN trained with high-density pilots (denoted as $F_H^*$) store informative knowledge of the channel. Using the parameters of $F_H^*$ as the initial parameters for training the DACEN with low-density pilots can transfer the knowledge embedded in the parameters. On the other hand, from the perspective of network training, generating neighboring low-density pilot samples from the high-density pilot samples for data augmentation can diversify original low-density pilot samples, thereby further improving the low-overhead DACEN-based channel estimation performance.

To reap the benefits from high-density pilots, in the model training stage, the BS transmits high-density pilots to the UEs, producing high-density pilot samples. Consider the source dataset $\mathcal{D}_H$ consisting of $N_{D_H}$ high-density pilot samples and the task of channel estimation with $\mathcal{D}_H$ as the source task $\mathcal{T}_H$. 
The spatial-temporal domain channel representations corresponding to the received pilot samples are used as the labels for training the DACEN. 
Denote $\underline{\mathbf{Y}}_{pH,i} \in \mathbb{C}^{N_R \times N_T \times N_H}$ as the $i$-th high-density pilot sample in $\mathcal{D}_H$ and $\underline{\mathbf{H}}_{t,i}$ as the corresponding channel label. The target dataset $\mathcal{D}_L$ contains low-density pilot samples, and the task of channel estimation with $\mathcal{D}_L$ is the target task $\mathcal{T}_L$. After the model training stage, the well-trained target model is deployed to the UEs. At this time, the BS transmits low-density pilots to the UEs, and then the UEs estimate the downlink channel with the deployed model given received low-density pilots.

With the proposed parameter-instance transfer learning approach, the model training process is shown in Fig. \ref{tf} (b). 
The proposed parameter-instance transfer learning approach includes three phases, i.e., the source model training phase, the sample generation and evaluation phase, and the target model training phase. In the source model training phase, we train the source model for channel estimation with high-density pilot samples. We anticipate that more complete knowledge can be learned and stored by model parameters. In the sample generation and evaluation phase, we sample low-density pilot samples from the high-density pilot samples by selecting adjacent pilot samples. This is then followed by the sample evaluation and exclusion in the obtained low-density pilot samples based on the cosine similarity criteria. The evaluation and exclusion processes are leveraged to avoid incorporating excessive dissimilar samples and successive negative transfer issues\cite{zhang2022survey}. In the target model training phase, we finally train the target model for low-overhead channel estimation with the generated low-density pilot samples and parameters transferred from the source model. The pseudo-code of the proposed transfer learning algorithm is given in Algorithm \ref{alg1}. Below we elaborate on the details of the proposed parameter-instance transfer learning approach according to Algorithm \ref{alg1}.

\textbf{Phase 1: Source model training (line 1 - line 5)}. Given the source dataset $\mathcal{D}_H$ which contains high-density pilot samples and corresponding channel labels $\{(\underline{\mathbf{Y}}_{pH,i},\underline{\mathbf{H}}_{t,i})|i=1,...,N_{D_H}\}$, we split it for training, validation, and testing purposes. With the training subset $\{(\underline{\mathbf{Y}}_{pH,i},\underline{\mathbf{H}}_{t,i})|i=1,...,N_{D_H}^{train}\}$, a DACEN $F_H$ is firstly trained for $\mathcal{T}_H$ by minimizing the MSE loss. Notice that $F_H$ is expected to achieve high estimation accuracy, and the optimal $F_H$ can be determined by (\ref{S_loss}) at the top of the next page, where $\|\cdot\|_2$ denotes L2 norm and $N_{D_H}^{train}$ is the number of samples for training. The parameters of the backbone network of the well-trained $F_H^*$ are then used as a good initialization to train $F_L$ for the low-overhead channel estimation $\mathcal{T}_L$.

\begin{figure*}
\begin{align} \label{S_loss}
F_H^* = \arg \min _{F_H} \frac{1}{N_{D_H}^{train}} \sum_{i=1}^{N_{D_H}^{train}} \|[\operatorname{Im}\{\underline{\mathbf{H}}_{t,i}\},\operatorname{Re}\{\underline{\mathbf{H}}_{t,i}\}] - F_H([\operatorname{Im}\{\underline{\mathbf{Y}}_{pH,i}\},\operatorname{Re}\{\underline{\mathbf{Y}}_{pH,i}\}])\|_2^2,
\end{align}
\hrulefill
\end{figure*}

\textbf{Phase 2: Sample generation and evaluation (line 6 - line 22). First, initialize an empty target dataset $\mathcal{D}_L$ to contain generated low-density pilot samples.} Then, select $N_L$ elements from $\underline{\mathbf{Y}}_{pH,i} \in \mathbb{C}^{N_R \times N_T \times N_H}$, i.e., $\{\underline{\mathbf{Y}}_{pH,i}^{(r)}|r=r_0, r_0+I, ..., r_0+I*(N_L-1)\}$, to generate a low-density pilot sample $\underline{\mathbf{Y}}_{pL1,i} \in \mathbb{C}^{N_R \times N_T \times N_L}$, where $r_0$ is the starting RB's index and $I$ is the index spacing between two neighboring elements. Note that each of the $N_L$ elements represents the received pilot signal in one RB. The generated samples and corresponding channel labels constitute the target dataset $\mathcal{D}_L$, which is also split into the training subset $\mathcal{D}_L^{train}$ (containing $N_{D_L}^{train}$ samples), validation subset $\mathcal{D}_L^{val}$, and testing subset $\mathcal{D}_L^{test}$, respectively. Specifically, $\mathcal{D}_L^{train}$ is copied to a new set $\mathcal{D}_{L_E}^{train}$ for containing extended samples to train $F_L$. Considering the coherence in frequency domain among neighboring RBs, two additional low-density pilot samples potentially correlated with $\underline{\mathbf{Y}}_{pL1,i}$ can be obtained as $\{\underline{\mathbf{Y}}_{pH,i}^{(r)}|r=r_0-1, r_0+I-1, ..., r_0+I*(N_L-1)-1\}$ and $\{\underline{\mathbf{Y}}_{pH,i}^{(r)}|r=r_0+1, r_0+I+1, ..., r_0+I*(N_L-1)+1\}$. Denote the two samples as $\underline{\mathbf{Y}}_{pL2,i} \in \mathbb{C}^{N_R \times N_T \times N_L}$ and $\underline{\mathbf{Y}}_{pL3,i} \in \mathbb{C}^{N_R \times N_T \times N_L}$. To avoid the negative impacts brought by incorporating uncorrelated data for training, the correlations among $\underline{\mathbf{Y}}_{pL1,i}$, $\underline{\mathbf{Y}}_{pL2,i}$ and $\underline{\mathbf{Y}}_{pL3,i}$ are also evaluated. Taking the similarity evaluation among $\underline{\mathbf{Y}}_{pL2,i}$ and $\underline{\mathbf{Y}}_{pL1,i}$ as an example, $\underline{\mathbf{Y}}_{pL2,i}$ and $\underline{\mathbf{Y}}_{pL1,i}$ are first rearranged to matrices $\mathbf{V}_{pL2,i} \in \mathbb{C}^{N_L\times (N_R \times N_T)}$ and $\mathbf{V}_{pL1,i} \in \mathbb{C}^{N_L\times (N_R \times N_T)}$, respectively. Then, the similarity can be evaluated by matching their $N_L$ row vector pairs one by one. Denote $\mathbf{a}_{12,i} \in \mathbb{R}^{N_L \times 1}$ the intermedia cosine similarity vector for all $N_L$ vector pairs, the $j$-th element of $\mathbf{a}_{12}$ representing the cosine similarity of the $j$-th vector pair $(\mathbf{V}_{pL2,i}^{(j)}, \mathbf{V}_{pL1,i}^{(j)})$ is:
\begin{equation} \label{s_j}
    \mathbf{a}_{12,i}^{(j)} = cossim(\mathbf{V}_{pL2,i}^{(j)},\mathbf{V}_{pL1,i}^{(j)}) = \frac{|<\mathbf{V}_{pL2,i}^{(j)}, \mathbf{V}_{pL1,i}^{(j)}>|}{\|\mathbf{V}_{pL2,i}^{(j)}\|_2 \|\mathbf{V}_{pL1,i}^{(j)}\|_2},
\end{equation}
where $cossim(\cdot)$ denotes the cosine similarity and $<\cdot, \cdot>$ denotes the inner product. Next, we average the cosine similarities of all $N_L$ vector pairs to get a final cosine similarity score:
\begin{equation} \label{s}
    \mathbf{s}_{12}^{(i)} = \frac{1}{N_L} \sum_{j=1}^{N_L} \mathbf{a}_{12,i}^{(j)},
\end{equation}
where $\mathbf{s}_{12}=[\mathbf{s}_{12}^1,\ldots,\mathbf{s}_{12}^{N_{D_L}^{train}}]^T \in \mathbb{R}^{N_{D_L}^{train} \times 1}$ is the cosine similarity vector for all $N_{D_L}^{train}$ sample pairs. The sample $\underline{\mathbf{Y}}_{pL2,i}$ is added to the extended training subset $\mathcal{D}_{L_E}^{train}$ once the final score $\mathbf{s}_{12}^{(i)}$ is larger or equal to a predefined threshold $s^{th}$ (e.g., $0.8$), and the final score is recorded as its instance weight. The sample generation process is operated on all high-density pilot samples, and the evaluation process is operated on all high-density pilot samples in the training subset $\mathcal{D}_{H}^{train}$.

\textbf{Phase 3: Target model training (line 23 - line 27)}. Given the generated low-density pilot samples $\{\underline{\mathbf{Y}}_{pL,i}|i=1,...,N_{D_{L_E}}^{train}\}$ (including $\underline{\mathbf{Y}}_{pL1,i}$, $\underline{\mathbf{Y}}_{pL2,i}$, or $\underline{\mathbf{Y}}_{pL3,i}$) in $\mathcal{D}_{L_E}^{train}$, corresponding channel labels $\{\underline{\mathbf{H}}_{t,i}|i=1,...,N_{D_L}^{train}\}$,\footnote{Note that $\underline{\mathbf{Y}}_{pL1,i}$, $\underline{\mathbf{Y}}_{pL2,i}$, and $\underline{\mathbf{Y}}_{pL3,i}$ share the same channel label $\underline{\mathbf{H}}_{t,i}$.} and the recorded instance weights $\mathbf{w}$, the proposed DACEN can then be trained based on the parameter initialization transferred from \textbf{Phase 1} and can be specified by (\ref{T_loss}) at the top of the next page, where $N_{D_{L_E}}^{train}$ is the number of samples in the extended training subset $\mathcal{D}_{L_E}^{train}$. Note that the instance weight for $\underline{\mathbf{Y}}_{pL1,i}$ is 1.

\begin{figure*}
\begin{align} \label{T_loss}
F_L^{TL} = \arg \min _{F_L} \frac{1}{N_{D_{L_E}}^{train}} \sum_{i=1}^{N_{D_{L_E}}^{train}} \mathbf{w}^{(i)}\|[\operatorname{Im}\{\underline{\mathbf{H}}_{t,i}\},\operatorname{Re}\{\underline{\mathbf{H}}_{t,i}\}] - F_L([\operatorname{Im}\{\underline{\mathbf{Y}}_{pL,i}\},\operatorname{Re}\{\underline{\mathbf{Y}}_{pL,i}\}])\|_2^2,
\end{align}
\hrulefill
\end{figure*}

It's worth noting that as shown in Fig. \ref{tf} (c), only low-density pilot samples are input to the trained DACENs to acquire the channel estimates in the model testing stage.

\begin{algorithm}
        \scriptsize
	\renewcommand{\algorithmicrequire}{\textbf{Input:}}
	\renewcommand{\algorithmicensure}{\textbf{Output:}}
        
	\caption{Parameter-instance transfer learning approach}
	\label{alg1}
	\begin{algorithmic}[1]
	\REQUIRE \ \\
	$\mathcal{D}_H$: source dataset, containing high-density pilot samples and corresponding channel labels, i.e., $\{(\underline{\mathbf{Y}}_{pH,i} ,\underline{\mathbf{H}}_{t,i})|i=1,...,N_{D_H}\}$ \\
 $\mathcal{T}_H$, $\mathcal{T}_L$: source task and target task \\
	$F_H$, $F_L$: DACENs for the source task $\mathcal{T}_H$ and the target task $\mathcal{T}_L$ \\
	$N^{itr}_H$, $N^{itr}_L$: maximum iterations for training $F_H$ and $F_L$ \\
	$N_L$: number of RBs in a low-density pilot sample\\
	$I$: distance to sample two neighboring elements from high-density pilot samples\\
	$r_0$: starting RB number of low-density pilots\\
	$s^{th}$: the cosine similarity threshold to include a sample
    \\ $\rule{\linewidth}{0.1mm}$
    \\ \textbf{Phase 1: Source model training}
	\STATE Initialization: split $\mathcal{D}_H$ into training, validation, testing subsets: $\mathcal{D}_H^{train}$, $\mathcal{D}_H^{val}$, $\mathcal{D}_H^{test}$; randomly initialize $F_H$; $m \leftarrow 0$
	\REPEAT
	\STATE Update $F_H$ according to (\ref{S_loss}) with samples in $\mathcal{D}_{H}^{train}$
	\STATE $m \leftarrow m + 1$
	\UNTIL $m \ge N^{itr}_H$
    \\ $\rule{\linewidth}{0.1mm}$
	\\ \textbf{Phase 2: Sample generation and evaluation}
	\STATE Initialization: initialize an empty target dataset $\mathcal{D}_L$ to contain generated low-density pilot samples \\
	\FOR{$i=1,...,N_H$}
	\STATE Sample $\{\underline{\mathbf{Y}}_{pH,i}^{(r)}|r=r_0, r_0+I, ..., r_0+I*(N_L-1)\}$ from $\underline{\mathbf{Y}}_{pH,i}$ to form a low-density pilot sample $\underline{\mathbf{Y}}_{pL1,i}$
	\STATE Insert $(\underline{\mathbf{Y}}_{pL1,i}$, $\underline{\mathbf{H}}_{t,i})$ into $\mathcal{D}_L$
	\ENDFOR
	\STATE Split $\mathcal{D}_L$ into training, validation, testing subsets: $\mathcal{D}_L^{train}$ (containing $N_{D_L}^{train}$ samples), $\mathcal{D}_L^{val}$, and $\mathcal{D}_L^{test}$
	\STATE $\mathcal{D}_{L_E}^{train} \leftarrow \mathcal{D}_L^{train}$
	\FOR{$i=1,...,N_{D_L}^{train}$}
	\STATE Sample $\{\underline{\mathbf{Y}}_{pH,i}^{(r)}|r=r_0-1, r_0+I-1, ..., r_0+I*(N_L-1)-1\}$ and $\{\underline{\mathbf{Y}}_{pH,i}^{(r)}|r=r_0+1, r_0+I+1, ..., r_0+I*(N_L-1)+1\}$ from $\underline{\mathbf{Y}}_{pH,i}$ to form $\underline{\mathbf{Y}}_{pL2,i}$ and $\underline{\mathbf{Y}}_{pL3,i}$
	\FOR{$k=2,3$}
        \STATE Rearrange $\underline{\mathbf{Y}}_{pLk,i}$ and $\underline{\mathbf{Y}}_{pL1,i}$ to $\mathbf{V}_{pLk,i}$ and $\mathbf{V}_{pL1,i}$
        \STATE Calculate the final cosine similarity score $\mathbf{s}_{1k}^{(i)}$ according to (\ref{s_j}) and (\ref{s})
	    \IF{$\mathbf{s}_{1k}^{(i)} \ge s^{th}$}
            \STATE Insert $(\underline{\mathbf{Y}}_{pLk,i}$, $\underline{\mathbf{H}}_{t,i})$ into $\mathcal{D}_{L_E}^{train}$
        \ENDIF
    \ENDFOR
    \ENDFOR
    \\ $\rule{\linewidth}{0.1mm}$
	\\ \textbf{Phase 3: Target model training}
	\STATE Initialization: initialize $F_L$ based on the optimal source model $F_H^*$, $m \leftarrow 0$
	\REPEAT
	\STATE Update $F_L$ according to (\ref{T_loss}) with samples in $\mathcal{D}_{L_E}^{train}$
	\STATE $m \leftarrow m + 1$
	\UNTIL $m \ge N^{itr}_L$
	\ENSURE \ \\
	optimal target model $F_L^{TL}$
	\end{algorithmic}
\end{algorithm}

\section{Experiments} \label{Sec. Exp.}
In this section, we first introduce experimental settings including system setup, hyper-parameter settings, the evaluation metric, etc, for the numerical evaluations. We then compare the performance of the proposed methods with existing methods under various pilot density settings. We further verify the effectiveness of the SAM and TAM through an ablation study.

\subsection{Experimental Settings}

To verify the effectiveness of the proposed DACEN-based low-overhead channel estimation method, we first construct a simulation dataset based on the clustered delay line (CDL) MIMO channel model with the MATLAB 5G toolbox \cite{3gpp2022tr38901,MATLAB5G}. A single-cell system served by one BS equipped with $32$ antennas is considered. The system adopts OFDM with $52$ RBs of the time-frequency resources, and every $12$ subcarriers and $14$ OFDM symbols constitute an RB. There are $2,000$ UEs randomly distributed in the single-cell system, of which $1,800$ UEs are used for training and validation and thus are configured to receive high-density pilots, while the rest $200$ UEs receive pilots with different SNR settings and pilot density settings to fully compare the proposed DACEN-based low-overhead channel estimation method with the baselines. Each UE is equipped with $4$ antennas and its velocity is $15$ km/h. The BS transmits CSI-RS signals over $26$ RBs to the UEs in the training and validation sets at the signal-to-noise ratio (SNR) of $10$ decibels (dB), and the received high-density pilots from ten evenly distributed slots among consecutive ten frames are recorded. As a result, $18,000$ high-density received pilot samples from $1,800$ UEs in the training and validation set with the pilot density of $\rho_H=26/52$ are provided for training ($16,200$ samples) and validation ($1,800$ samples). While for testing, given a specific SNR and pilot density setting, $2,000$ received pilot samples from $200$ UEs in the testing set are used for performance evaluation. The SNR settings, pilot density settings, and other key system parameters are listed in Table \ref{SimulationSetup}.

\begin{figure}[htbp]
\centering
\includegraphics[width=0.45\textwidth]{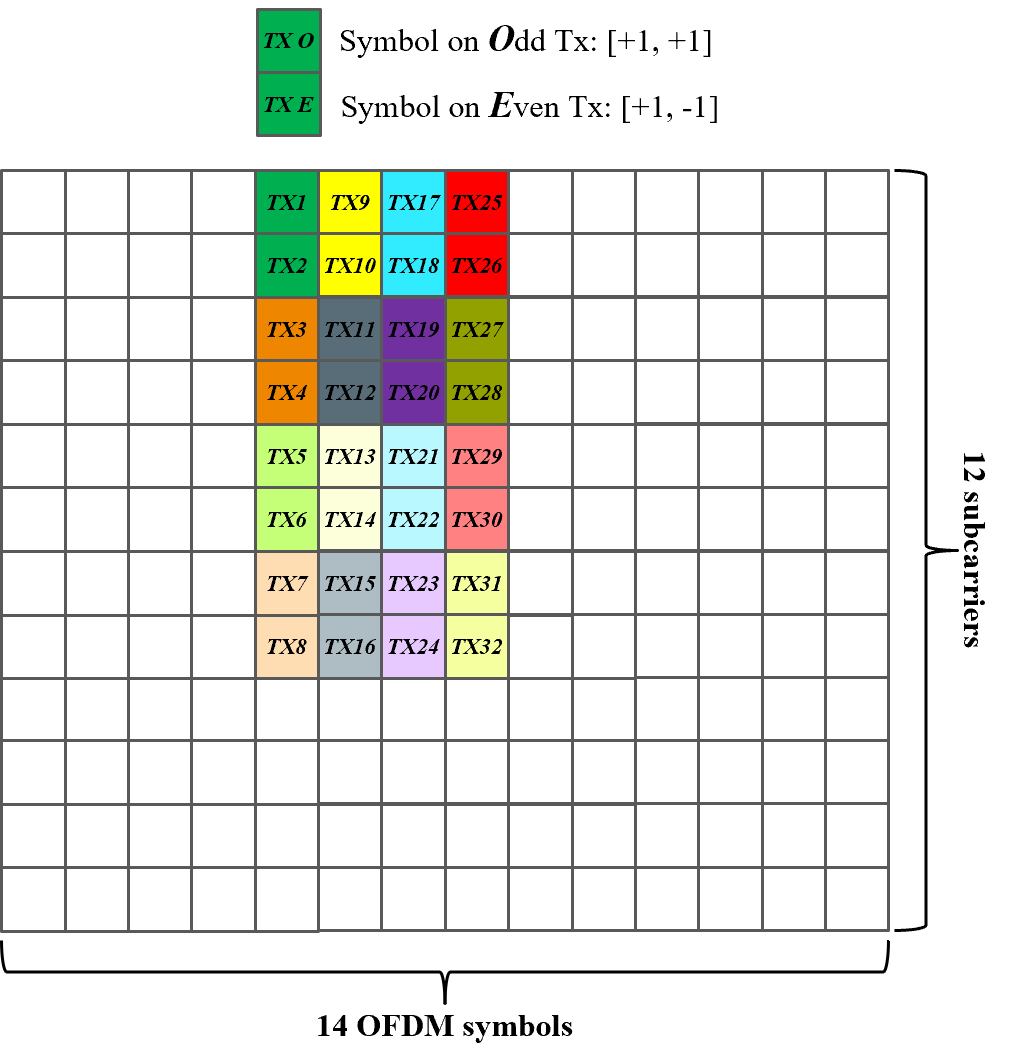}
\caption{The CSI-RS configuration within one RB.}
\label{SystemConfig}
\end{figure}

The configuration of CSI-RS within one RB for the channel measurement among multiple antenna ports is designed as shown in Fig. \ref{SystemConfig}. $32$ resource elements are divided into 16 groups, and within each group, only two TX ports transmit reference signals (i.e., [$+1$, $+1$] and [$+1$, $-1$], respectively). The details of the CSI-RS configuration are referred to \cite{3gpp2023ts38211}.

Note that the measured spatial-frequency domain channel representation corresponds to $624$ frequency domain subcarriers. With an IFFT operation as (\ref{IFFT}) shows ($N_P = 64$), the spatial-temporal channel representation has only $64$ entries in the temporal domain channel response, which is about only $1/10$ the number of subcarriers of a spatial-frequency channel representation. Therefore, taking the spatial-temporal channel representation as the label for training the DNN-based channel estimator can reduce tremendous memory usage and model complexity.

\begin{table}[htbp]
\centering
\caption{System parameter settings.}
\label{SimulationSetup}
\resizebox{\columnwidth}{!}{%
\begin{tabular}{c|c}
\hline
\textbf{Parameter}           & \textbf{Value}                           \\ \hline
Channel model                & CDL                                      \\ \hline
Number of clusters\footnotemark           & $14$                                     \\ \hline
Rician K-factor (dB) in the first cluster          & $-10$                                    \\ \hline
Channel sampling rate        & $61.44$ MHz                              \\ \hline
FFT size                     & $1024$                                   \\ \hline
Carrier frequency            & $3.5$ GHz                                \\ \hline
Subcarrier spacing           & $60$ kHz                                 \\ \hline
UE velocity                  & $15$ km/h                                \\ \hline
$N_{RB}$                     & $52$                                     \\ \hline
$N_T$                        & $32$                                     \\ \hline
$N_R$                        & $4$                                      \\ \hline
$N_c$                        & $624$                                    \\ \hline
The number of UEs in the training and validation set & $1,800$           \\ \hline
The number of UEs in the testing set & $200$                            \\ \hline
SNR settings (dB)            & $5$, $10$, $15$, $20$, $25$         \\ \hline
Pilot density settings       & $26/52$, $6/52$, $4/52$, $2/52$ \\ \hline
\end{tabular}
}
\end{table}

\footnotetext{The delay profile of the clusters is defined the same as the CDL-E channel model\cite{3gpp2022tr38901}.}

\begin{table}[htbp]
\centering
\caption{Final hyper-parameter settings for the proposed DACEN.}
\label{Hyper}
\begin{tabular}{c|c}
\hline
\textbf{Hyper-parameter}                        & \textbf{Setting} \\ \hline
\textbf{representation dimension   $d_\text{model}$}     & $512$              \\ \hline
\textbf{feed forward dimension   $d_\text{ff}$} & $512$              \\ \hline
\textbf{attention head $N_h$}                   & $2$                \\ \hline
\textbf{no. of SA modules $N_{SA}$}              & $8$                \\ \hline
\textbf{no. of TA modules $N_{TA}$}              & $8$                \\ \hline
\textbf{PE hyper-parameter $\omega$}              & $10000$ \\ \hline
\textbf{batch size $N_{bs}$}                        & $256$              \\ \hline
\textbf{initial learning rate $\eta$}             & $6e-5$             \\ \hline
\end{tabular}
\end{table}

To improve the generality of the proposed DACEN, data samples collected from distributed UEs with diverse scenarios in the training set are utilized to train the model. This is based on the following observations. First, neighboring UEs share some scatterers, and UEs in different locations may have the same patterns of movement. Modeling these physical correlations by using data from multiple locations is meaningful to improve the generality of deep learning approaches. Second, the training process is a process in which the model learns to map received pilots to channel matrices. Utilizing training data from distributed UEs with diverse scenarios would probably ease the learning process and lead to a model with better generality.

The DACEN is implemented with PyTorch\cite{paszke2019pytorch}, and we use the Adam optimizer \cite{kingma2014adam} to train the DACEN with two NVIDIA 3090 GPU cards. The hyper-parameter settings of the proposed DACEN-based method are shown in Table \ref{Hyper}. The normalized mean squared error (NMSE) in dB is adopted to evaluate the channel estimation performance, which is defined as:
\begin{equation}\label{NMSE}
    \text{NMSE (dB)} =  10 * \log \left(\frac{1}{N}\sum_{i=1}^{N} \frac{\|\underline{\mathbf{H}}_{t,i} - \hat{\underline{\mathbf{H}}}_{t,i}\|_2^2}{\|\underline{\mathbf{H}}_{t,i}\|_2^2}\right),
\end{equation}
where $N$ is the number of samples.

We also evaluate the computational complexity of the proposed DACEN in terms of the time complexity and the space complexity by floating-point multiply-accumulate operations (`FLOPs') and the number of parameters (`Params'), respectively.

\subsection{Effectiveness of the Proposed DACEN-based Channel Estimation Methods}

The proposed DACEN-based channel estimation method is compared with several baselines and state-of-the-art (SOTA) methods for channel estimation under various SNRs and pilot density settings. Specifically, we compare the proposed DACEN-based method with the traditional least-square (LS) estimation-based method \cite{li2000optimum} and linear minimum mean-squared error (LMMSE) estimation-based method \cite{choi2015lowcomplexity}. For the LS-based channel estimation method, the traditional LS algorithm is first exploited on subcarriers with known pilot symbols, followed by an interpolation method to obtain the estimate of all subcarriers. While for the LMMSE-based method, two Wiener filtering matrices are obtained to exploit the correlations among adjacent antennas and among adjacent subcarriers, respectively. Then the LMMSE estimate is achieved based on the LS estimate and \cite{li2000optimum}. In addition to traditional methods, the performance comparison between the proposed method and deep learning-based channel estimation methods, including the FC-DNN-based channel estimation method\cite{ye2018power}, the CNN-based channel estimation method \cite{dong2019deep}, the CDRN\cite{liu2022deep}, and the skip-connection attention (SC-attention) network\cite{liu2023deep}, is also presented. Note that we also tune the hyper-parameters of the deep learning-based baselines to achieve their best performance for fair comparisons.

\begin{figure}[htbp]
\centering
\begin{minipage}[t]{0.48\textwidth}
\centering
\includegraphics[width=8cm]{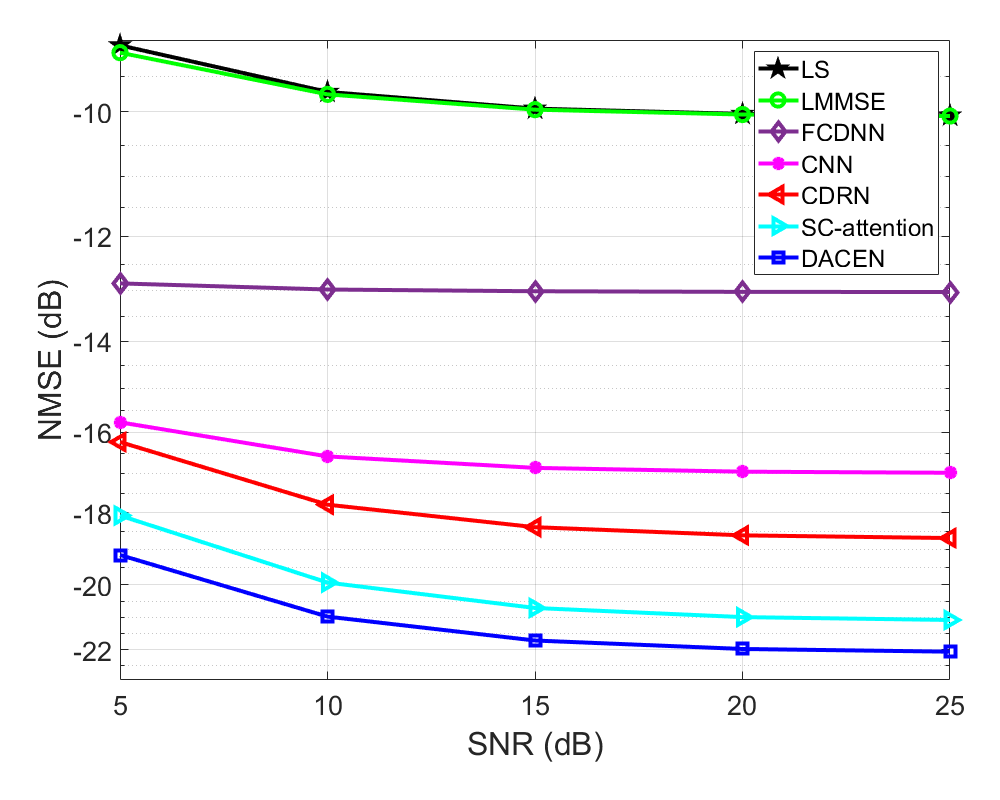}
\caption{NMSE performance under different SNRs with $\rho_H=26/52$.}
\label{26RB}
\end{minipage}
\begin{minipage}[t]{0.48\textwidth}
\centering
\includegraphics[width=8cm]{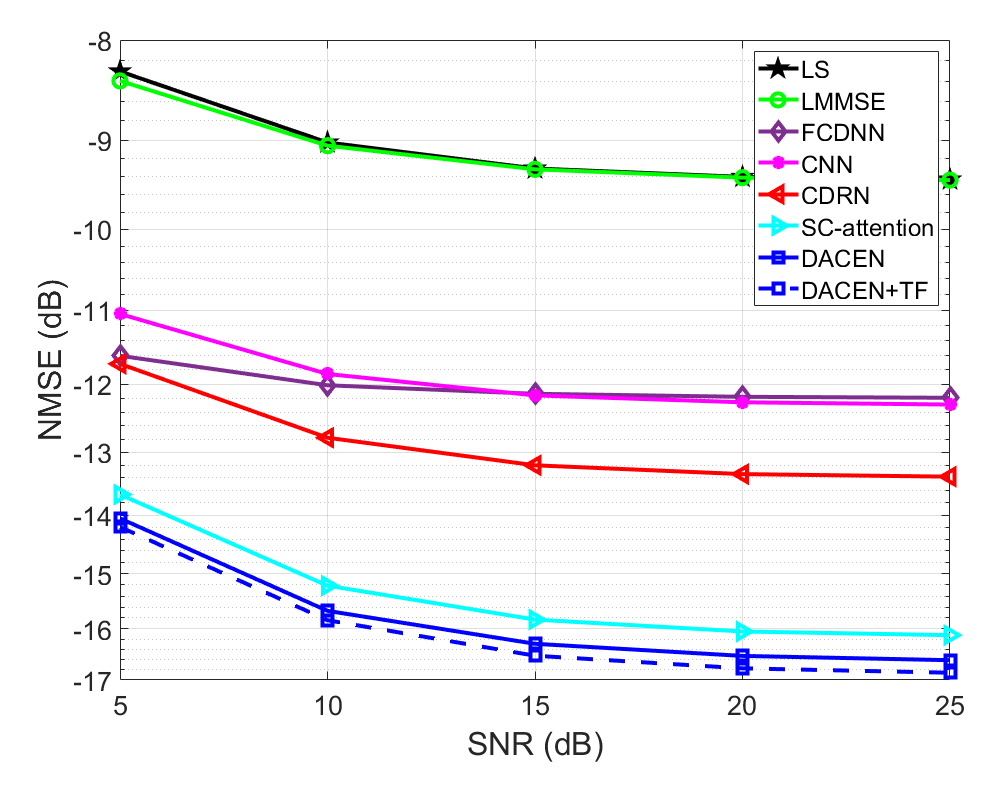}
\caption{NMSE performance under different SNRs with $\rho_L=6/52$.}
\label{6RB}
\end{minipage}
\end{figure}

\begin{figure}[htbp]
\centering
\begin{minipage}[t]{0.48\textwidth}
\centering
\includegraphics[width=8cm]{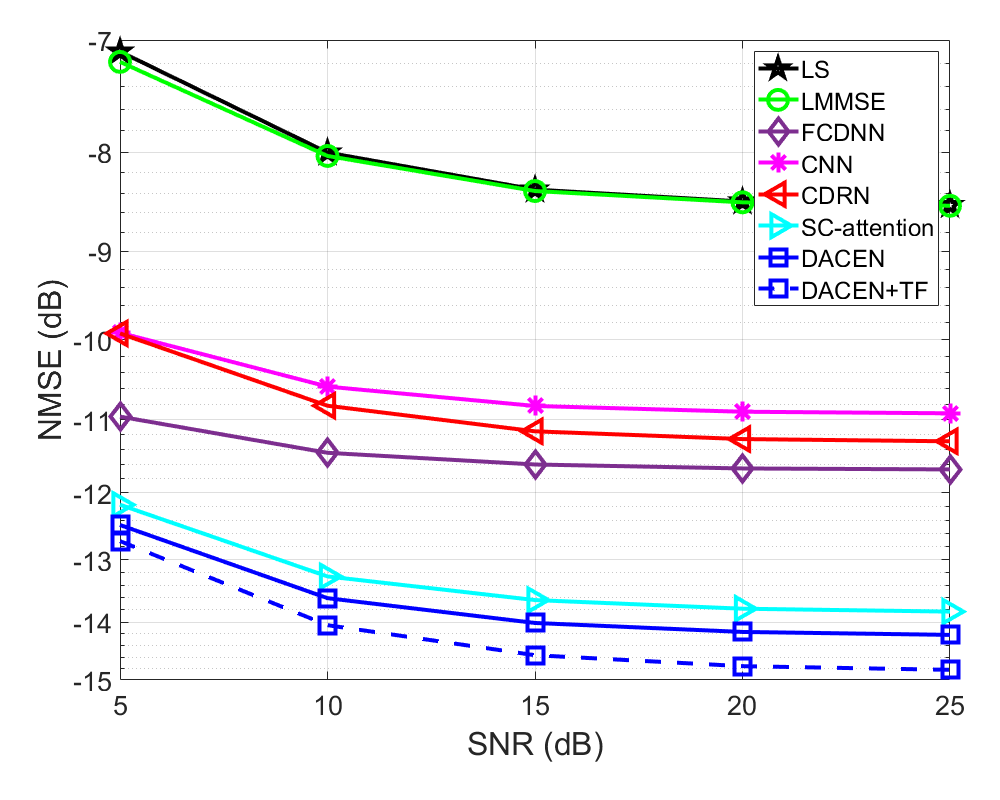}
\caption{NMSE performance under different SNRs with $\rho_L=4/52$.}
\label{4RB}
\end{minipage}
\begin{minipage}[t]{0.48\textwidth}
\centering
\includegraphics[width=8cm]{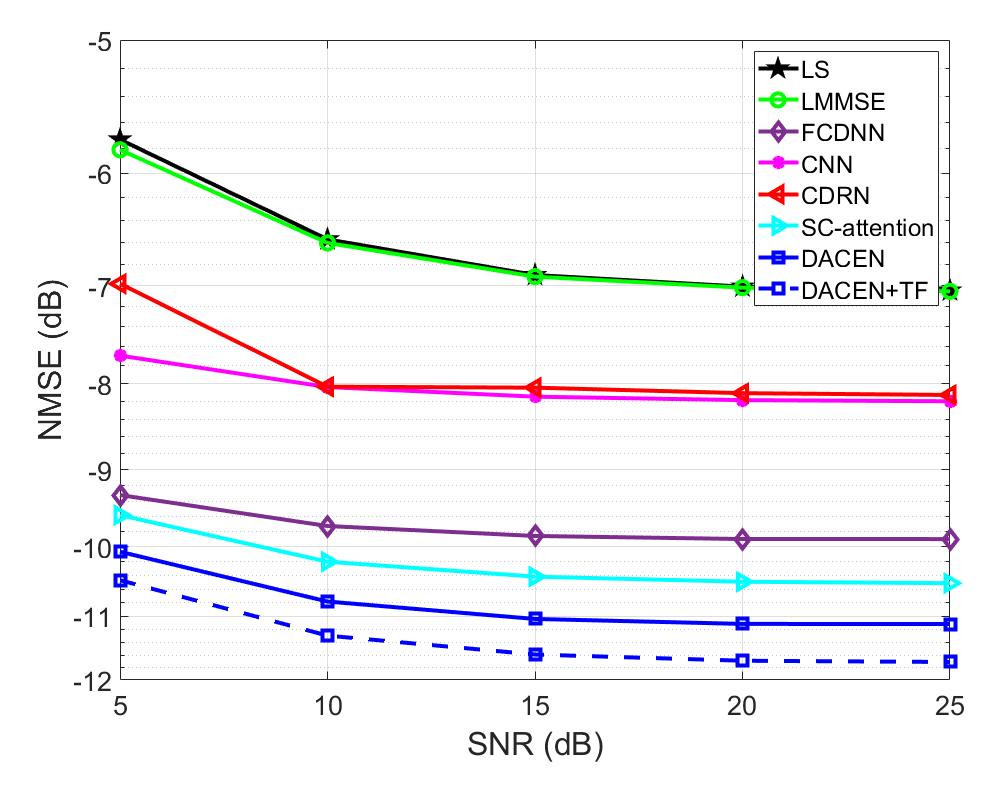}
\caption{NMSE performance under different SNRs with $\rho_L=2/52$.}
\label{2RB}
\end{minipage}
\end{figure}

The results of NMSE under different SNRs and pilot density settings are presented in Fig. \ref{26RB} - Fig. \ref{2RB}. Note that the DACEN (denoted by `\textbf{DACEN}') in Fig. \ref{6RB} - Fig. \ref{2RB} is trained from scratch with low-density pilots, while the `\textbf{DACEN+TF}' is trained with the proposed transfer learning algorithm following Algorithm \ref{alg1}. To better compare the proposed method with other deep learning-based methods, their computational complexity is shown in Table \ref{model_complexity}.\footnote{Table \ref{model_complexity} represents the model complexity when the inputs are low-density pilots with $\rho_L=6/52$ during the testing stage.} It can be seen from these results that for a certain pilot density, the proposed DACEN-based method outperforms all other methods under all SNRs. Specifically, the proposed DACEN-based method with $\rho_L=2/52$ achieves better performance than traditional LS and LMMSE methods with high-density pilots. And the performance of the proposed DACEN-based method with $\rho_L=6/52$ also shows competitive performance as the FCDNN and CNN estimators with $\rho_H=26/52$. This, consequently, further corroborates the capability of the proposed DACEN in the pilot overhead reduction. When compared to SOTA methods, i.e., the CDRN and the SC-attention network, the proposed DACEN-based generally achieves $3$ dB and $1$ dB advantages of NMSE performance with high-density pilots, and $3$ dB and $0.6$ dB advantages of NMSE performance with $\rho_L=2/52$. In addition, from Table \ref{model_complexity}, the DACEN has much lower computational complexity than the CDRN and the SC-attention network, demonstrating the high efficiency of the proposed DACEN-based method.

The NMSE performance of `\textbf{DACEN+TF}' is achieved under three low-density pilot settings, i.e., $\rho_L=6/52$, $\rho_L=4/52$, and $\rho_L=2/52$. In correspondence with Algorithm \ref{alg1}, we set $r_0=1$, $I=4$, and $N_L=6$ for $\rho_L=6/52$, $r_0=5$, $I=5$, and $N_L=4$ for $\rho_L=4/52$, and $r_0=9$, $I=8$, and $N_L=2$ for $\rho_L=2/52$, respectively. The cosine similarity threshold $s^{th}$ is set to $0.9$. The DACEN benefits from high-density pilots with $\rho_H=26/52$ by knowledge transfer during the training process, thereby obtaining up to $0.25$ dB, $0.6$ dB, and $0.6$ dB additional performance improvement than training from scratch. As presented in Table \ref{model_complexity}, it is worth emphasizing that the proposed transfer learning algorithm does not increase model complexity during the testing stage, therefore it will not bring additional burden to the UE.

Note that the proposed method can be directly applied to the single-cell multi-user scenarios once the proposed DACEN is deployed at multiple users. The CSI-RS can be set in a broadcast mode by the base station to all users in the cell so that all users can receive and demodulate it. In this case, there is no interference between different users due to the downlink broadcast of CSI-RS. Therefore, although our simulation results are based on a single-user scenario, the results characterize MU-MIMO performance to some extent.

\begin{table}[htbp]
\centering
\caption{Computational complexity comparison of the proposed method and other deep learning-based methods.}
\label{model_complexity}
\resizebox{\columnwidth}{!}{%
\begin{tabular}{c|cc}
\hline
\multirow{2}{*}{\textbf{Method}} & \multicolumn{2}{c}{\textbf{Computational Complexity}}                                                   \\ \cline{2-3} 
                                 & \multicolumn{1}{c|}{\textbf{Time Complexity (FLOPs)}} & \textbf{Space Complexity (Params)} \\ \hline
\textbf{FC-DNN}                  & \multicolumn{1}{c|}{$19.00$ M}                        & $18.90$ M                          \\ \hline
\textbf{CNN}                     & \multicolumn{1}{c|}{$2.50$ G}                         & $19.52$ M                          \\ \hline
\textbf{CDRN}                    & \multicolumn{1}{c|}{$3.40$ G}                         & $26.58$ M                          \\ \hline
\textbf{SC-attention}            & \multicolumn{1}{c|}{$3.74$ G}                         & $29.21$ M                          \\ \hline
\textbf{DACEN}                   & \multicolumn{1}{c|}{$2.31$ G}                         & $17.01$ M                          \\ \hline
\textbf{DACEN+TF}                & \multicolumn{1}{c|}{$2.31$ G}                         & $17.01$ M                          \\ \hline
\end{tabular}
}
\end{table}

\subsection{Ablation Study}
 
\begin{figure}[htbp]
\centering
\begin{minipage}[t]{0.48\textwidth}
\centering
\includegraphics[width=8cm]{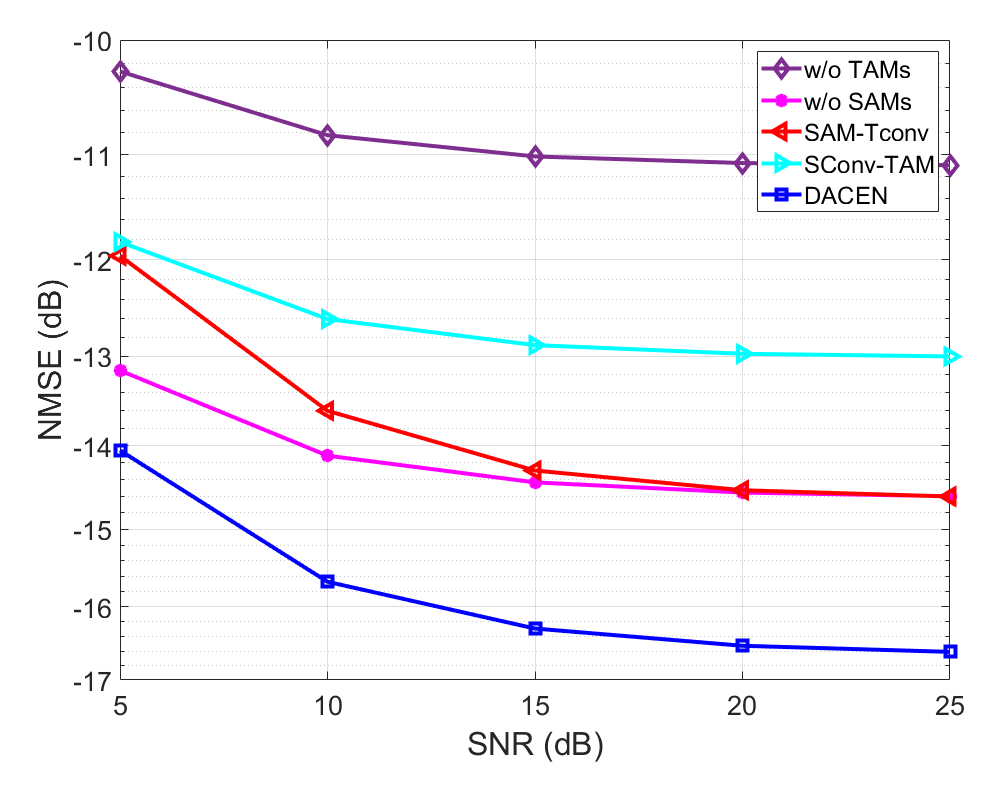}
\caption{NMSE performance of modified DNN models under different SNRs with $\rho_L=6/52$.}
\label{ablation}
\end{minipage}
\end{figure}

To further evaluate the effectiveness and computational efficiency of the TAM and the SAM, we conduct an ablation study. Given a fixed pilot density $\rho_L=6/52$, we first respectively remove all SAMs and all TAMs from the DACEN. In addition, we replace the SAMs in the DACEN with 2-dimensional convolutional layers (i.e., spatial convolutional layers (SConvs) with filter size $K_S \times K_S = 3 \times 3$). Moreover, we replace the TAMs in the DACEN with 1-dimensional convolutional layers (i.e., temporal convolutional layers (TConvs) with filter size $K_T = 3$). For simplicity, the obtained four modified DNN models are denoted as `\textbf{w/o SAMs}', `\textbf{w/o TAMs}', `\textbf{SConv-TAM}', and `\textbf{SAM-TConv}', respectively. The experiment results are shown in Fig. \ref{ablation} and the computational complexity analysis of these modified DNN models is presented in Table \ref{ablation_complexity}. In addition, the per-layer computational complexity of the TAM, TConv, SAM, and SConv is shown in Table \ref{per_layer}. 

From Fig. \ref{ablation}, removing all SAMs (`\textbf{w/o SAMs}') or all TAMs (`\textbf{w/o TAMs}') leads to obvious performance degradation, indicating the necessity and importance of the deployment of the TAM and the SAM. By replacing the SAMs with the SConvs or replacing the TAMs with the TConvs, the `\textbf{SConv-TAM}' and the `\textbf{SAM-TConv}' also perform worse than the proposed DACEN. This further emphasizes the effectiveness of the designed SAM and TAM.

According to Table \ref{ablation_complexity} and Table \ref{per_layer}, although the computational complexity of the TAM is nearly two times as the TConv, the TAM can learn the temporal dependencies of different delayed paths and brings significant performance improvement (i.e., around $2$ dB gain on average, comparing the DACEN with `\textbf{SAM-TConv}'). The SAM also achieves better estimation performance than the SConv (i.e., around $3$ dB gain on average, comparing the DACEN with `\textbf{SConv-TAM}') with only around $1/5$ the computational complexity of the SConv, thereby demonstrating the effectiveness and computational efficiency of the SAM.

\begin{table}[htbp]
\centering
\caption{Computational complexity comparison of modified DNN models for ablation study.}
\label{ablation_complexity}
\resizebox{\columnwidth}{!}{%
\begin{tabular}{c|cc}
\hline
\multirow{2}{*}{\textbf{Method}} & \multicolumn{2}{c}{\textbf{Computational Complexity}}                                                   \\ \cline{2-3} 
                                 & \multicolumn{1}{c|}{\textbf{Time Complexity (FLOPs)}} & \textbf{Space Complexity (Params)} \\ \hline
\textbf{DACEN}                   & \multicolumn{1}{c|}{$2.31$ G}                           & $17.01$ M                            \\ \hline
\textbf{TAM}                     & \multicolumn{1}{c|}{$1.76$ G}                           & $12.74$ M                            \\ \hline
\textbf{SAM}                     & \multicolumn{1}{c|}{$0.55$ G}                           & $4.27$ M                             \\ \hline
\textbf{SConv-TAM}               & \multicolumn{1}{c|}{$3.96$ G}                           & $29.92$ M                            \\ \hline
\textbf{SAM-TConv}               & \multicolumn{1}{c|}{$1.35$ G}                           & $10.57$ M                            \\ \hline
\end{tabular}
}
\end{table}

\begin{table}[htbp]
\centering
\caption{Per-layer computational complexity of the TAM, TConv, SAM, and SConv.}
\label{per_layer}
\resizebox{\columnwidth}{!}{%
\begin{tabular}{c|cc}
\hline
\multirow{2}{*}{\textbf{Method}} & \multicolumn{2}{c}{\textbf{Computational Complexity}}                                                   \\ \cline{2-3} 
                                 & \multicolumn{1}{c|}{\textbf{Time Complexity (FLOPs)}} & \textbf{Space Complexity (Params)} \\ \hline
\textbf{TAM}                     & \multicolumn{1}{c|}{$218.47$ M}                         & $1.58$ M                             \\ \hline
\textbf{TConv}                   & \multicolumn{1}{c|}{$100.73$ M}                         & $786.94$ k                           \\ \hline
\textbf{SAM}                     & \multicolumn{1}{c|}{$67.18$ M}                          & $525.31$ k                           \\ \hline
\textbf{SConv}                   & \multicolumn{1}{c|}{$302.06$ M}                         & $2.36$ M                             \\ \hline
\end{tabular}
}
\end{table}

\section{Conclusion} \label{Sec. Con.}
This paper proposed the DACEN-based channel estimation method for massive MIMO systems with low-density pilots. The DACEN comprises two attention modules, i.e., the TAM and the SAM, to jointly learn the spatial-temporal domain features of massive MIMO channels. The TAM is designed to learn the temporal dependencies among different delayed paths, while the SAM is designed to learn the correlation of CSI among different antennas. By exploiting the spatial correlation of massive MIMO channels, the SAM has much lower time and space complexity when compared to convolutional neural networks. This is significantly valuable for massive MIMO systems, especially when a larger and larger antenna array is deployed. Moreover, to further improve the channel estimation accuracy, we also proposed a parameter-instance transfer learning approach based on the DACEN to transfer the channel knowledge learned from the high-density pilots during the model training phase to that with low-density pilots. Experimental results based on various pilot density settings revealed the effectiveness of the proposed joint spatial-temporal domain feature extraction network and the superiority of the proposed DACEN-based low-overhead channel estimation method.

\section{Acknowledgement}
We acknowledge IMT-2020(5G), CAICT, and OPPO for raising the research problem of reducing pilot density in current 5G systems. This work was performed in part at SICC which is supported by SKL-IOTSC, University of Macau.

\bibliographystyle{IEEEtran}
\bibliography{refer.bib}

\begin{IEEEbiography}
    [{\includegraphics[width=1in,height=1.25in,clip,keepaspectratio]{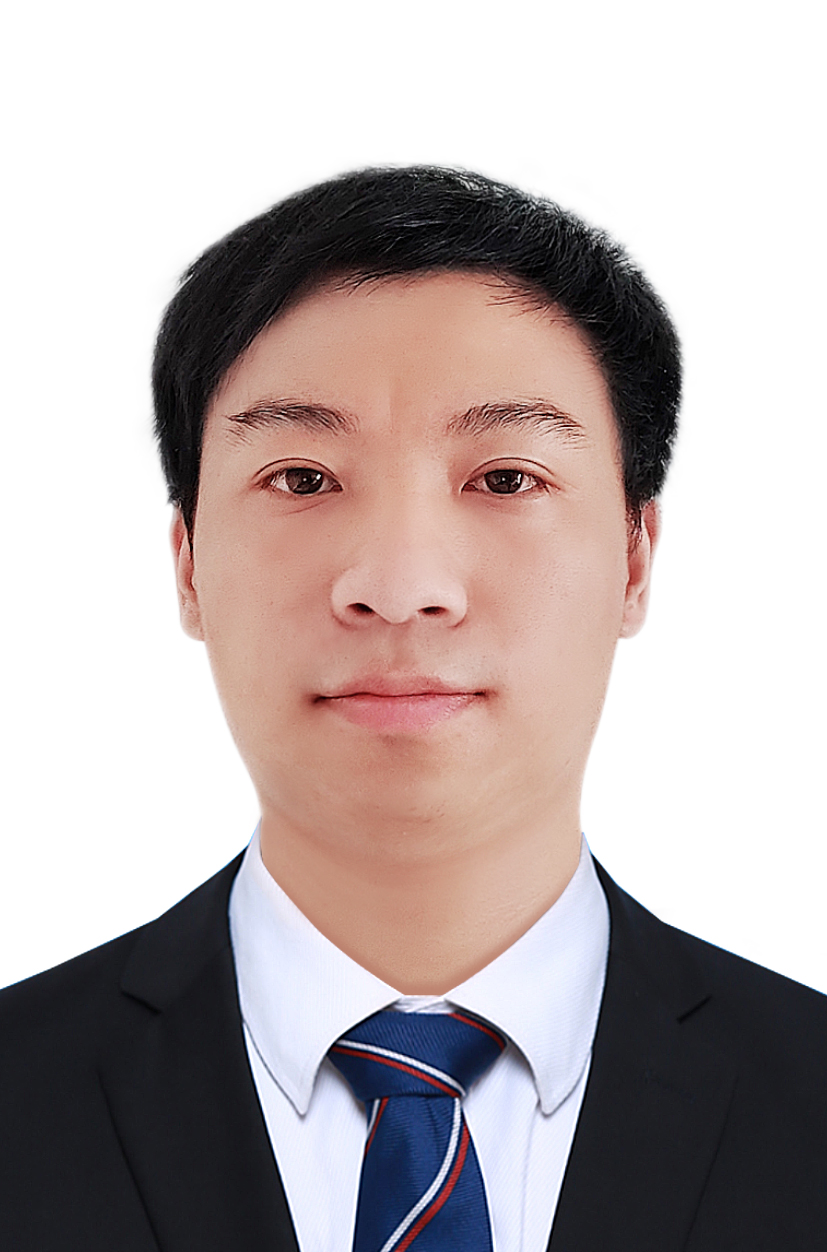}}]{Binggui Zhou} received the B.Eng. degree from Jinan University, Zhuhai, China, in 2018, and the M.Sc. degree from the University of Macau, Macao, China, in 2021, respectively. He is currently working toward the Ph.D. degree in Electrical and Computer Engineering with the University of Macau, Macao, China. He also serves as a Research Assistant with the School of Intelligent Systems Science and Engineering, Jinan University, Zhuhai, China. His research interests include artificial intelligence, intelligent wireless communications, and spatial-temporal data mining.
\end{IEEEbiography}

\begin{IEEEbiography}
    [{\includegraphics[width=1in,height=1.25in,clip,keepaspectratio]{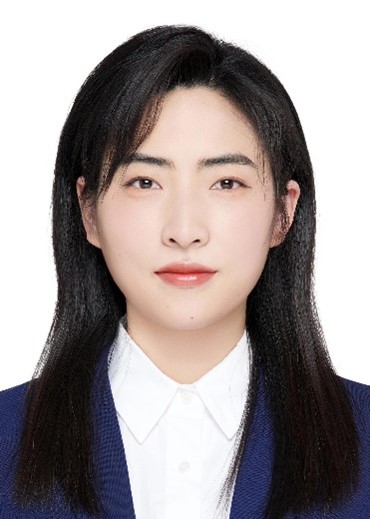}}]{Xi Yang} received the B.S., M.Eng. and Ph.D. degrees from Southeast University, Nanjing, China, in 2013, 2016 and 2019, respectively. From July 2020 to July 2022, she was a postdoctoral fellow with the State Key Laboratory of Internet of Things for Smart City, University of Macau, China. She is currently a Zijiang Young Scholar with the School of Communication and Electronic Engineering, East China Normal University, Shanghai, China. Her current research interests include extremely large aperture array (ELAA) systems, millimeter wave communications, and wireless communication system prototyping.
\end{IEEEbiography}

\begin{IEEEbiography}
    [{\includegraphics[width=1in,height=1.25in,clip,keepaspectratio]{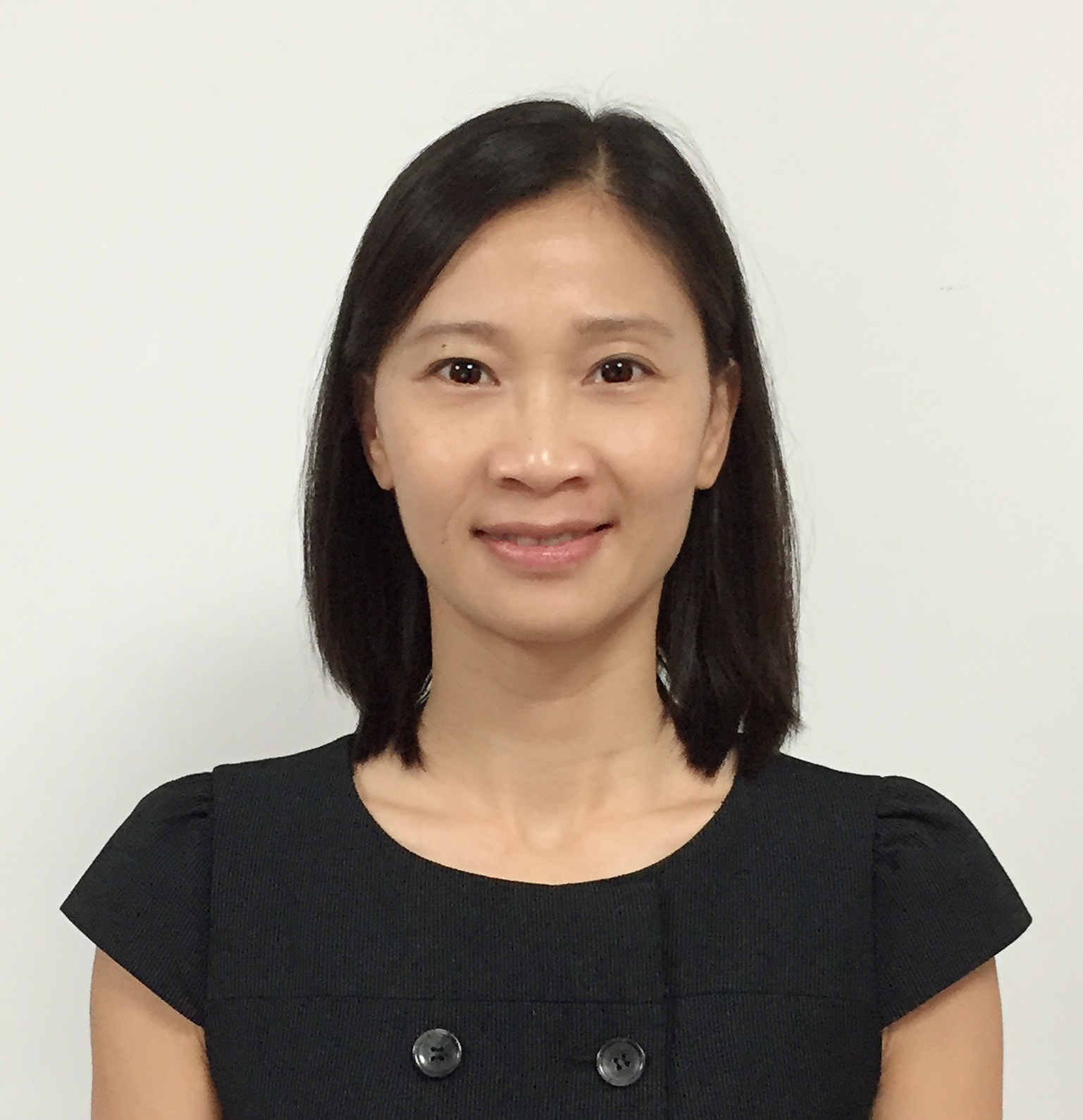}}]{Shaodan Ma} received the double Bachelor's degrees in science and economics and the M.Eng. degree in electronic engineering from Nankai University, Tianjin, China, in 1999 and 2002, respectively, and the Ph.D. degree in electrical and electronic engineering from The University of Hong Kong, Hong Kong, in 2006. From 2006 to 2011, she was a post-doctoral fellow at The University of Hong Kong. Since August 2011, she has been with the University of Macau, where she is currently a Professor. Her research interests include array signal processing, transceiver design, localization, integrated sensing and communication, mmwave communications and massive MIMO. She was a symposium co-chair for various conferences including IEEE ICC 2021, 2019 \& 2016, IEEE/CIC ICCC 2019, IEEE GLOBECOM 2016, etc. Currently she serves as an Editor for IEEE Transactions on Wireless Communications, IEEE Transactions on Communications, IEEE Communications Letters, and Journal of Communications and Information Networks.
\end{IEEEbiography}

\begin{IEEEbiography}
    [{\includegraphics[width=1in,height=1.25in,clip,keepaspectratio]{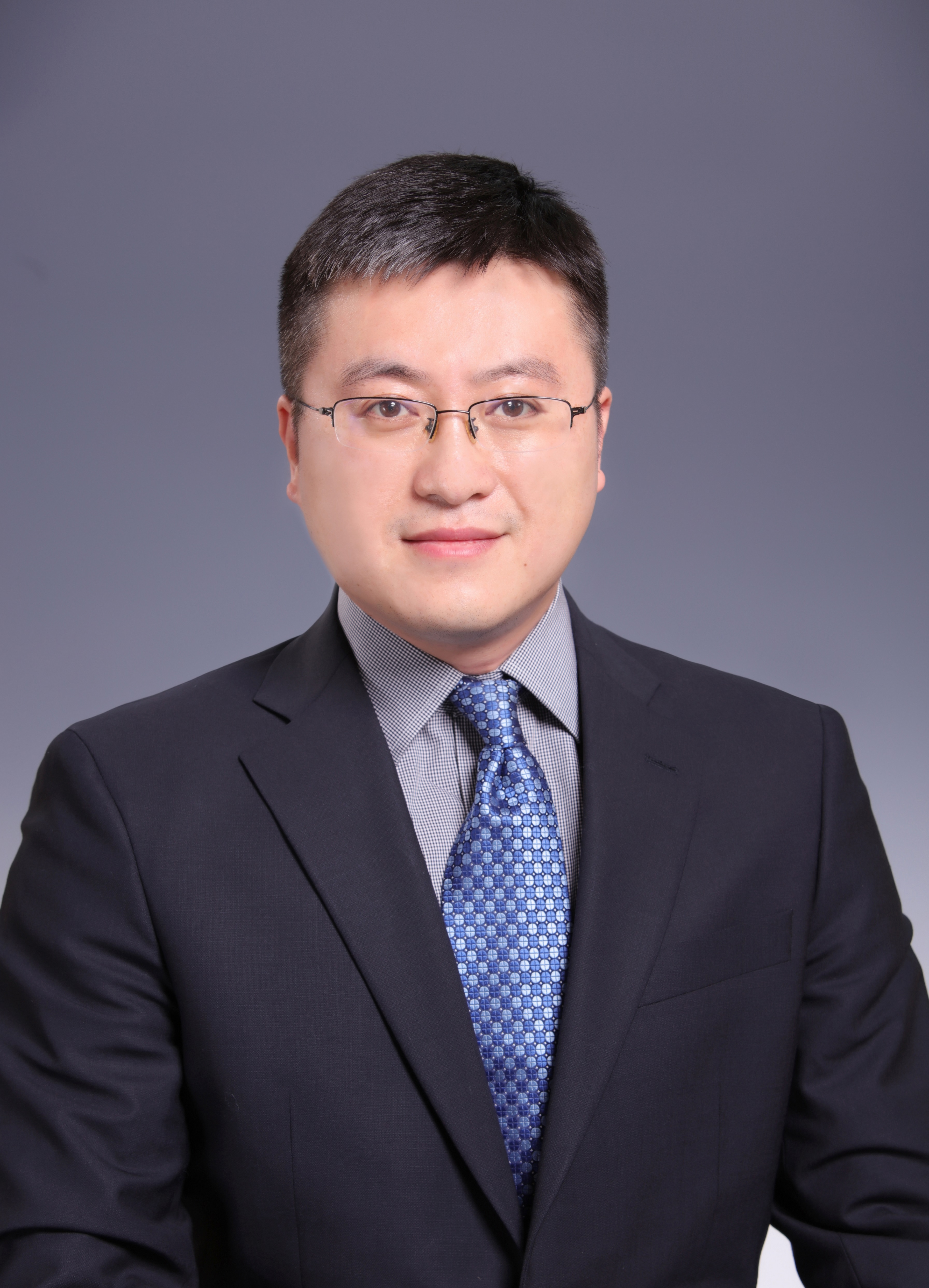}}] {Feifei Gao} (M'09-SM'14-F'20) received the B.Eng. degree from Xi'an Jiaotong University, Xi'an, China in 2002, the M.Sc. degree from McMaster University, Hamilton, ON, Canada in 2004, and the Ph.D. degree from National University of Singapore, Singapore in 2007. Since 2011, he joined the Department of Automation, Tsinghua University, Beijing, China, where he is currently an Associate Professor. 
    
    Prof. Gao's research interests include signal processing for communications, array signal processing, and artificial intelligence assisted communications. He has authored/coauthored more than 150 refereed IEEE journal papers and more than 150 IEEE conference proceeding papers that are cited more than 16000 times in Google Scholar. Prof. Gao has served as an Editor of IEEE Transactions on Communications, IEEE Transactions on Wireless Communications, IEEE Journal of Selected Topics in Signal Processing (Lead Guest Editor), IEEE Transactions on Cognitive Communications and Networking, IEEE Signal Processing Letters (Senior Editor), IEEE Communications Letters (Area Editor), IEEE Wireless Communications Letters, and China Communications. He has also served as the symposium co-chair for 2019 IEEE Conference on Communications (ICC), 2018 IEEE Vehicular Technology Conference Spring (VTC), 2015 IEEE Conference on Communications (ICC), 2014 IEEE Global Communications Conference (GLOBECOM), 2014 IEEE Vehicular Technology Conference Fall (VTC), as well as Technical Committee Members for more than 50 IEEE conferences.
\end{IEEEbiography}

\begin{IEEEbiography}
    [{\includegraphics[width=1in,height=1.25in,clip,keepaspectratio]{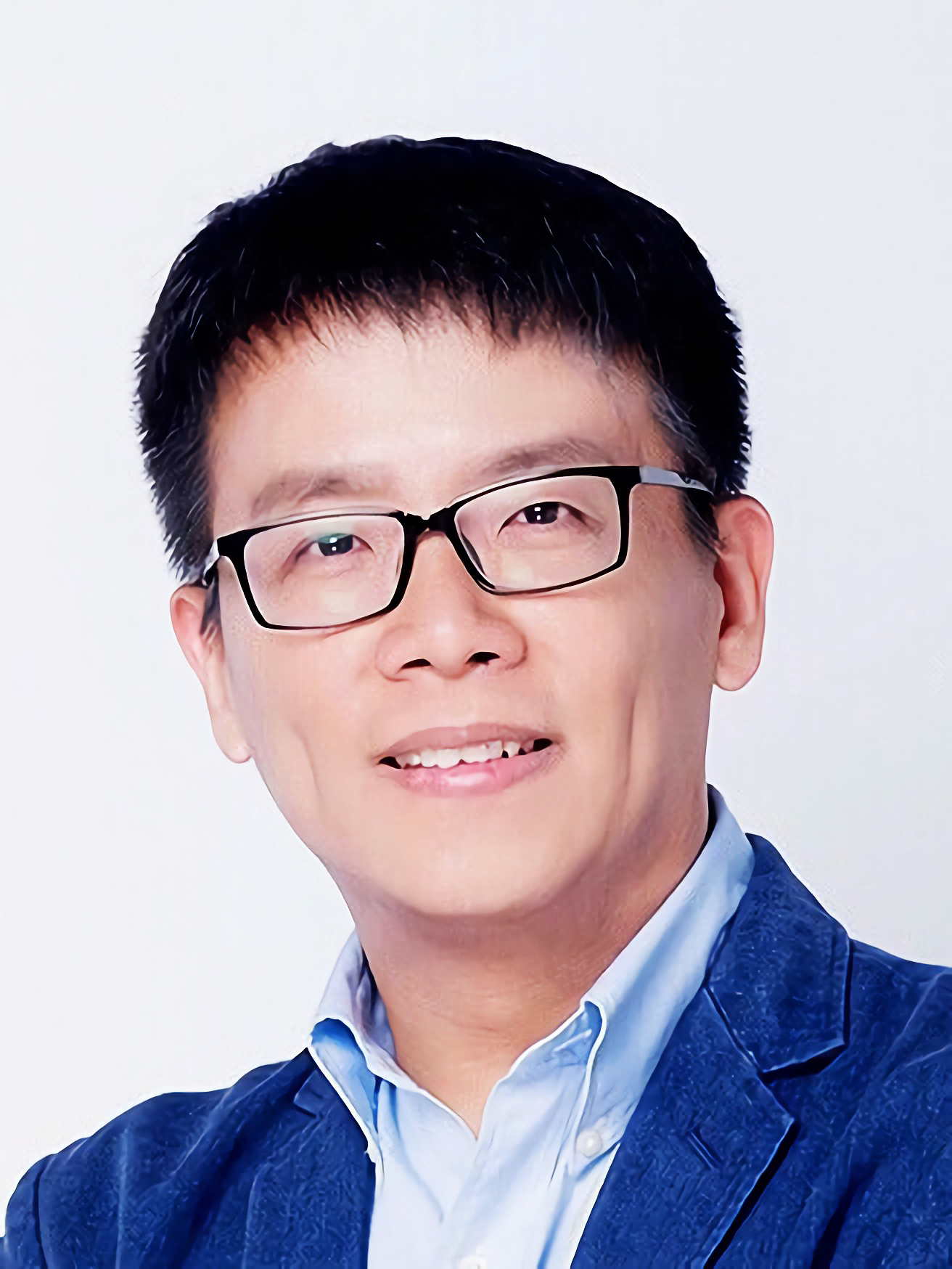}}]{Guanghua Yang} received his Ph.D. degree in electrical and electronic engineering from the University of Hong Kong in 2006. From 2006 to 2013, he served as post-doctoral fellow, research associate at the University of Hong Kong. Since April 2017, he has been with Jinan University, where he is currently a Full Professor in the School of Intelligent Systems Science and Engineering. His research interests are in the general areas of AI and communications.
\end{IEEEbiography}

\end{document}